\documentclass[aps,prb,superscriptaddress,nofootinbib,twocolumn,amsmath,amssymb]{revtex4-2}
\usepackage[colorlinks,linkcolor=red,anchorcolor=blue,citecolor=blue]{hyperref}
\usepackage{graphicx} 
\usepackage{amsmath}
\usepackage{amssymb}
\usepackage{bbm}
\usepackage{bm}
\usepackage{multirow}
\usepackage{mathtools}
\usepackage{subcaption}
\usepackage{array}
\usepackage[dvipsnames]{xcolor}
\usepackage{ragged2e}
\usepackage[font=small,labelfont=bf,justification=justified]{caption}
\usepackage{dsfont}
\usepackage[export]{adjustbox}
\usepackage{soul}

\begin{document}
\title{Attention is all you need to solve chiral superconductivity}
\author{Chun-Tse Li}
\thanks{These three authors contributed equally.}
\affiliation{Institute of Physics, Academia Sinica, Taipei 115201, Taiwan}
\affiliation{Department of Electrical and Computer Engineering, University of Southern California, Los Angeles, California 90089, USA}
\author{Tzen Ong}
\thanks{These three authors contributed equally.}
\affiliation{Institute of Physics, Academia Sinica, Taipei 115201, Taiwan}
\author{Max Geier}
\thanks{These three authors contributed equally.}
\affiliation{Department of Physics, Massachusetts Institute of Technology, Cambridge, MA 02139, USA}
\author{Hsin Lin}
\affiliation{Institute of Physics, Academia Sinica, Taipei 115201, Taiwan}
\author{Liang Fu}
\affiliation{Department of Physics, Massachusetts Institute of Technology, Cambridge, MA 02139, USA}

\newcommand{\R}{\mathbb{R}}
\newcommand{\C}{\mathbb{C}}
\newcommand{\Z}{\mathbb{Z}}

\begin{abstract}
    Recent advances on neural quantum states have shown that correlations between quantum particles can be efficiently captured by {\it attention} -- a foundation of modern neural architectures that enables neural networks to learn the relation between objects.  
    In this work, we show that a general-purpose self-attention Fermi neural network is able to find chiral $p_x \pm i p_y$ superconductivity in an attractive Fermi gas by energy minimization, {\it without prior knowledge or bias towards pairing}.  
    The superconducting state is identified from the optimized wavefunction by measuring various physical observables.
    %
    %
   We develop a symmetry projection method that reveals the ground state angular momentum and time-reversal symmetry breaking, 
    and a computation of the full two-body reduced density matrix spectrum that reveals the off-diagonal long-range order due to the dominant chiral $p$-wave pairing channel.
    Our work paves the way for AI-driven discovery of unconventional and topological superconductivity in strongly correlated quantum materials. 
    %
\end{abstract}


\maketitle

Solving the ground state of quantum many‑body systems is a central problem in condensed matter physics, as it underlies our understanding of quantum materials and their rich phase diagrams~\cite{hohenberg1964inhomogeneous,kohn1965self,
ceperley1980ground,foulkes2001quantum,becca2017quantum,white1992density,verstraete2023density}. 
A host of recent experiments have uncovered unconventional superconductivity and strongly correlated states, including fractional Chern insulators (FCIs) and chiral $d$-wave superconductivity, in multi-layer systems spanning graphene \cite{YacobyScience2010,NovoselovScience2011,cao2018insulator,Cao2018Apr,Yankowitz2019superconductivity,zhou2021halfmetal,zhou2022superconductivity-bernal,lu2024fqaheGraphene,lu2025qahe}, transition metal dichalcogenides and cuprates \cite{xu2020correlated,KimScience2023,tang2020hubbard,cai2023fqaheMoTe2,redekop2024fci,Xia2025Jan,Guo2025Jan}; in particular, signatures of chiral superconductivity have been observed in rhombohedral graphene \cite{Han2025Chiral}. These experimental successes demand the development of numerical techniques that accurately solve for the phase diagram of these systems, while sufficiently flexible to handle different experimental configurations.

Concurrently, advances in machine learning have opened a new path to solving quantum many-body problems variationally by using deep neural networks~\cite{carleo2017solving,pfau2020ab,hermann2020deep,von2022self} to universally approximate continuous functions~\cite{cybenko1989approximation, funahashi1989approximate, hornik1989multilayer}, including ground state wavefunctions of bosons and fermions. Neural network (NN) wavefunctions containing a large number of parameters can be efficiently optimized by energy minimization in a variational Monte Carlo (VMC) framework. This approach has proven to be successful in a number of fermion systems, especially for continuous space Hamiltonians~\cite{Pescia2024MessagePassing,luo2024simulatingmoire,XLi2025Moire,Smith2024Unified}. 




Despite rapid progress, existing research on neural network variational Monte Carlo has largely employed problem-specific neural quantum states (NQS). As an example, for spin-$\frac{1}{2}$ Fermi gas with repulsive interaction, a determinant based NN wavefunction is used to study the Fermi liquid ground state, whereas for the attractive case a paired wavefunction is introduced to study superconductivity~\cite{lou2024neural,Kim2024UltraCold,luo2023pairing}. It is therefore unclear whether there exists a universal neural network architecture applicable across a wide range of quantum systems. 
Only recently has a unifying neural network architecture based on the self-attention mechanism been proposed, tested on a variety of many-body systems, and shown to succeed {\it without pre-training or prior knowledge} \cite{geier2025attention, teng2025solving}.

In this work, we leverage the self-attention neural network to solve the problem of spin-polarized (or spinless) two-dimensional Fermi gas with attractive interaction, and find a superconducting ground state with chiral $p_x \pm i p_y$ pairing, which {\it spontaneously} breaks time-reversal symmetry (TRS).  We show that this chiral superconductor is topologically nontrivial over a wide range of interaction strengths, as evidenced by a peculiar ``odd-even'' effect distinct from conventional superconductors. 
%

Starting from first principles, our NN model faithfully captures the effect of quantum fluctuations beyond the BCS mean-field theory~\cite{read2000paired} and accurately solves the ground state of a {\it strong-coupling} chiral superconductor. 
In contrast to previous studies on superconductivity with neural network wavefunctions \cite{lou2024neural,kim2024neural}, our results demonstrate that a {\it self-attention} neural network wavefunction requires no pairing-specific modifications to describe the superconducting state.
Thereby, our achievement establishes that the same self-attention wavefunction thus solves molecules \cite{von2022self}, Wigner crystallization \cite{geier2025attention}, fractionalization \cite{teng2025solving} and now chiral superconductivity -- from first principles and without any modifications specific to the anticipated ground state structure. 
%

Our results showcase the ability of the NN to train the generalized Slater determinant wavefunction to the correct degenerate superconducting manifold strictly using energy minimization. 
We invent a new symmetry-projection method that reveals the chiral pairing: In a chiral $p_x \pm i p_y$ superconducting state, each Cooper pair contributes angular momentum $l_z = \pm 1$, and 
this fingerprint is revealed by comparing energies of the different angular momentum-projected NQS wavefunction. Furthermore, we fully characterize the off-diagonal long-range order (ODLRO) and chiral pairing structure \cite{YangRMP1962} via a spectral decomposition of the full two-body reduced density matrix (2-RDM) for the continuum many-body wavefunction.




{\it System.}--- We consider spin-polarized fermions with attractive Gaussian interaction in two dimensions,
\begin{align}
    H=-\frac{1}{2}\sum_{i=1}^N\nabla_i^2 + \frac{U}{2\pi\sigma_U^2}\sum_{i>j} \exp\left(-\frac{|\mathbf r_i-\mathbf r_j|^2}{2\sigma_U^2}\right)
\end{align}
where $U<0$ sets the coupling strength between fermions and the parameter $\sigma_U$ sets the interaction range.

{\em Neural Network Variational Monte Carlo.---}
Our wavefunction ansatz is constructed by enforcing only the most fundamental physical requirement of fermionic many-particle wavefunctions: the Pauli principle.
This antisymmetric structure under particle exchange is captured by a sum of determinants of {\it many-body} orbitals \cite{pfau2020ab}  
\begin{equation}
\Psi(\mathbf X)
=\frac{1}{\sqrt{N!}} \sum_{k=1}^{N_{\rm det}}
\det\bigl[\Phi^k_\mu(\mathbf x_j;\{\mathbf x_{/j}\})\bigr]_{j,\mu=1}^N.
\label{eq:wf_ansatz}
\end{equation}
where $\Phi_\mu\bigl(\mathbf x_j;\{\mathbf x_{/j}\}\bigr)$ depend on all particle coordinates and are permutation invariant under particle coordinates $\{ \mathbf{x}_{/j}\} := \{\mathbf x_1,\dots,\mathbf x_N\}\setminus \mathbf x_j$ except $\mathbf{x}_j$. 
Inspired by backflow transformation \cite{Cohen1956Backflow,KwonPRB1993,LuoPRL2019Backflow},  
the many-body fermion wavefunction Eq.~\eqref{eq:wf_ansatz} has been employed in neural network simulations of molecules \cite{Hermann2019PauliNet,pfau2020ab,von2022self,Gao2023Jul,Hermann2023Review,scherbela2024towards,Li2024ForwardLaplacian,foster2025abinitiofoundationmodel}, lattice models \cite{viteritti2023transformer,gu2025solvinghubbardmodelneural}, and continuum solids \cite{Li2022Solids,Cassella2023Jan,Wilson2023Jun,gerard2024transferableSolids,Pescia2024MessagePassing,geier2025attention}. 
mportantly, unlike Pfaffian or geminal type ansatz \cite{lou2024neural,kim2024neural}, the determinant based wavefunction Eq.~\eqref{eq:wf_ansatz} has no pairing structure built in. Remarkably, our self-attention network driven solely by energy minimization nonetheless finds chiral superconductivity.



\begin{figure}
    \centering
    \includegraphics[width=0.45\linewidth]{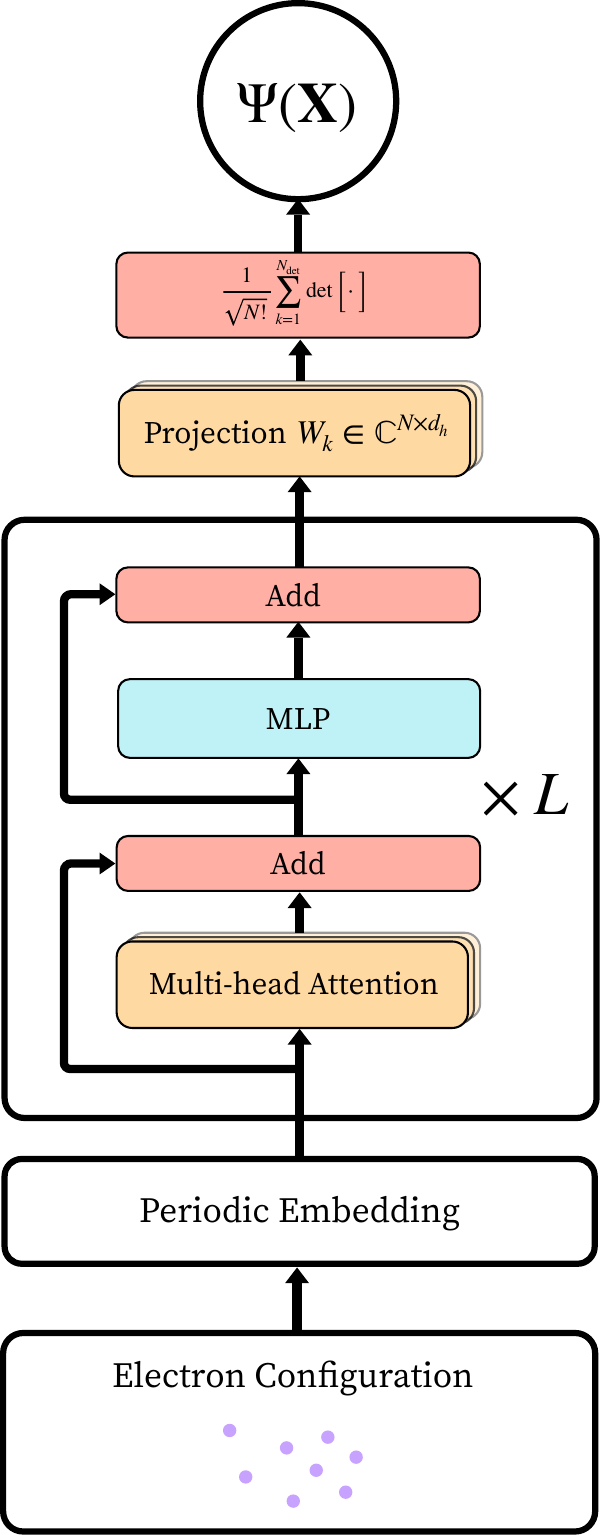}
    \caption{\justifying Neural quantum state architecture. Particle coordinates $\mathbf X$ sampled by an MCMC routine are embedded with periodic functions and then passed through permutation-equivariant self-attention layers. The network outputs generalized orbitals that form orbital matrices $\Phi^{(k)}$; the wavefunction is represented as a sum of generalized Slater determinants, $\Psi(\mathbf X)=\sum_k \tfrac{1}{\sqrt{N!}}\,\det[\Phi^{(k)}(\mathbf X)]$.}
    \label{fig:Psiformer-Architecture}
\end{figure}

{\em Neural network ansatz.---}
We consider the system within a supercell specified by the vectors $\mathbf L_j$ with periodic boundary condition. A transformer neural network \cite{vaswani2017attention} is employed to generate the many-body orbitals $\Phi_\mu\bigl(\mathbf x_j;\{\mathbf x_{/j}\}\bigr)$, see Fig.~\ref{fig:Psiformer-Architecture} for an architecture overview. 
Our architecture is identical to Ref.~\cite{geier2025attention} and follows the spirit of Ref.~\cite{von2022self}. We briefly outline the workflow below and refer to Ref.~\cite{geier2025attention} and App.~\ref{app:NN-Architecture} for details on the network architecture and App.~\ref{app:VMC} for the VMC technique. The hyperparameters used are listed in SM Table~\ref{table:hyperparameters}, and our results are robust and consistent across a range of different hyperparameters (see SM Fig.~\ref{fig:convergence-analysis}). Numerical calculations in this paper build on the recently developed code ``PeriodicWave'', which is publicly available \cite{periodicwave_github}.

To satisfy supercell periodicity, the network inputs are particle coordinates specified by sine and cosines of $\mathbf G_n^T \mathbf x_j$ where $\mathbf G_n^T$ are the primitive reciprocal supercell vectors satisfying $\mathbf G_n^T \mathbf L_m=2\pi \delta_{nm}$. Each particle input is then mapped to a corresponding vector in a high-dimensional vector space $\bm h_j^{(0)} \in \mathbb{R}^{d_{\rm int}}$  through  a linear transformation. 
These ``particle tokens'' $\mathbf h^{(0)}_j$ are processed by $L$ layers of multi-head self-attention layer followed by multi-layer perceptron (MLP) layers. 
The self-attention mechanism passes information between particle tokens $\mathbf h^{(l)}_j$,  thus introducing electron correlations while preserving the permutation equivariance. 
Finally, the network's output vectors $\{ \mathbf h^{(L)}_j \}$ are projected to $ N_{\det}$ sets of $N$ complex numbers that correspond to the many-body orbitals $\Phi_\mu\bigl(\mathbf x_j;\{\mathbf x_{/j}\}\bigr)$, from which the many-body wavefunction Eq.~\eqref{eq:wf_ansatz} is obtained for given particle coordinates $\{\mathbf x_j\}$.

Notice that all the neural‐network parameters are initialized randomly and optimized \emph{from scratch}, without  pretraining and without warm-starts from mean-field (HF/BCS), or problem-specific orbitals~\cite{li2022ab,scherbela2024towards,rende2024fine}. This ensures the learned chiral state emerges without built-in pairing structure or human bias.

{\it Pair-binding energy---} The presence of electron pairing is diagnosed by the pair binding energy,
\begin{align}
\label{eq:binding_energy}
    E_B(N) &= E(N) + E(N+2) - 2\,E(N+1),
\end{align}
where $E(N)$ is the ground‑state energy of the $N$‑particle sector. In spin-singlet superconductors with full gap, the ground state prefers to have an even number of electrons because of pairing. 
On the other hand, a hallmark of topological superconductivity in a spinless Fermi gas is a reversed even–odd effect~\cite{read2000paired,fu2010electron}. While pairing takes place between opposite momenta states $\pm \mathbf k$, there exists an “unpaired’’ state $\mathbf k=0$ lying below the Fermi level. This makes odd‑$N$ sectors energetically favored, leading to $E_B<0$ for odd $N$ and $E_B>0$ for even $N$. This is opposite to both spin-singlet superconductors and the strongly-paired spinless superconductors, where charge-$2e$ electron molecules form a BEC, which always prefers an even number of electrons and is topologically trivial.      

\begin{figure}[t]
    \centering
    \begin{subfigure}[b]{0.49\textwidth}
         \includegraphics[width=1.03\linewidth, right]{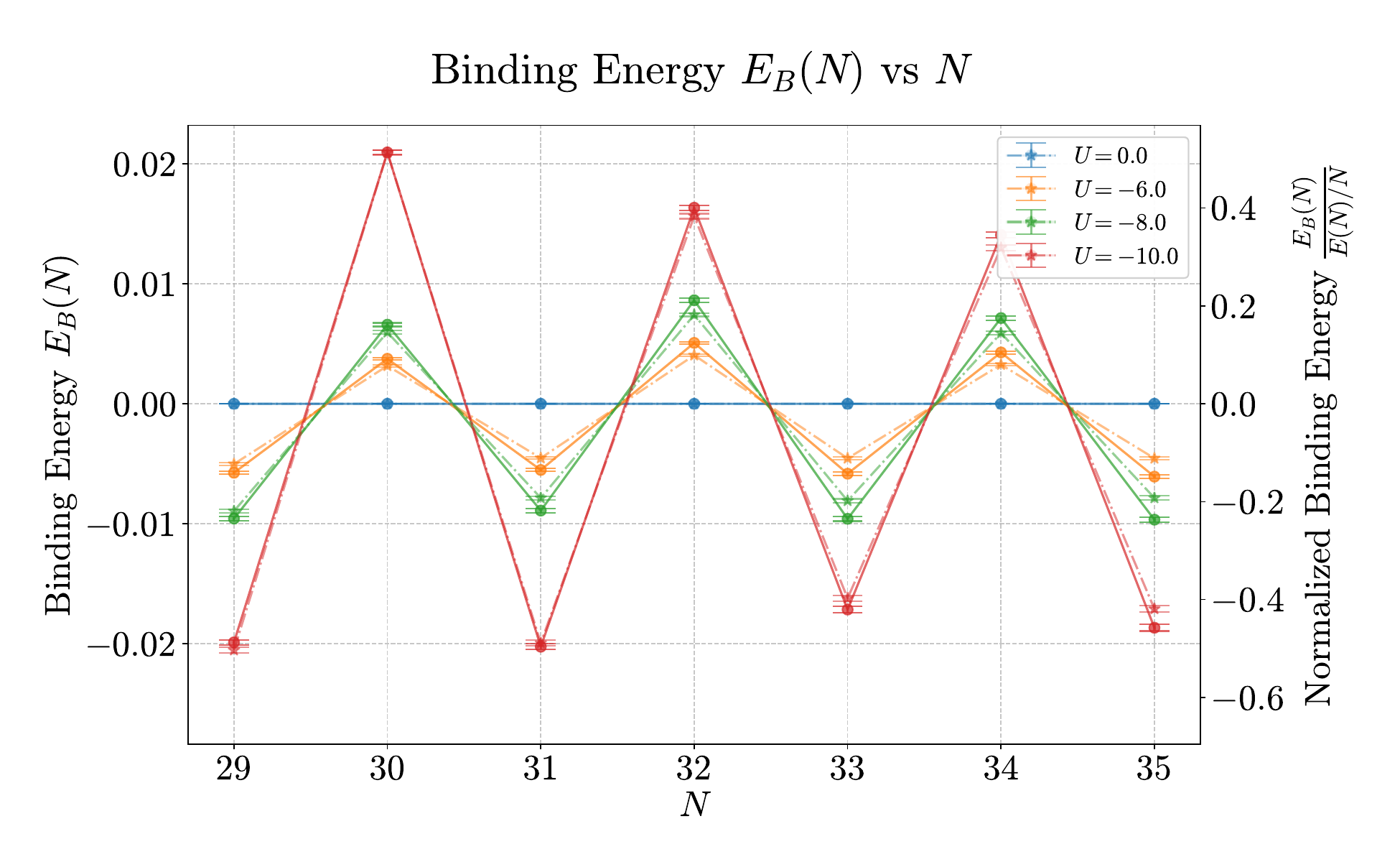}
    \end{subfigure}
    \caption{\justifying Even–odd effect in the pair‑binding energy $E_B(N)$ [Eq.~\eqref{eq:binding_energy}] versus particle number $N$ for $U=0,-6,-8,-10$ (solid lines); the dash–dotted lines show the normalized quantity $E_B(N)/(E(N)/N)$.  Error bars are one–standard–error Monte‑Carlo uncertainties.  For $U=0$ the degenerate Fermi surface gives $E_B(N)=0$. A clear even–odd oscillation appears for $U < 0$: odd-$N$ sectors 
    have negative $E_B$, while even-$N$ sectors 
    are positive -- consistent with a single unpaired $\mathbf{k}=0$ state and Cooper pairing of the remaining $2n$ particles.  
    }
    \label{fig:pair-binding}
\end{figure}

In Fig.~\ref{fig:pair-binding} we plot the pair‑binding energy obtained from our NN energies for various particle numbers.  
 We compute $E_B(N)$ for $N=29,\dots,37$, where $N=29$ and $N=37$ correspond to closed‑shell configurations in noninteracting limit $U=0$.  
 For all attraction strengths studied here, the pairing energy is finite and $|E_B|$ grows rapidly with $U$. Moreover, the sign of $E_B(N)$ shows the energetic preference of odd‑$N$ ground states, which provides first evidence for topological superconductivity. Note that for large $U$, the pair binding energy reaches as large as $40\%$ of the average energy per particle, indicating a strong-coupling superconductor.   

{\em Ground-state angular momentum of $p_x\pm ip_y$ superconductors---} Unlike $s$-wave superconductors, a $p_x \pm ip_y$ superconductor hosts Cooper pairs that each carry orbital angular momentum $\pm 1$. In the topologically nontrivial ground state of attractive Fermi gas, all but one electron at $\mathbf k=0$ pair up, resulting in a total angular momentum $M=\pm (N-1)/2$ ($N$ is odd). The sign depends on the chirality of the superconducting order, which occurs spontaneously 
in thermodynamic limit 
and breaks time reversal symmetry.      

On the finite square torus studied here, the continuous rotation symmetry is reduced to $C_4$ symmetry, hence the ground states are eigenstates of $C_4$ rotation, with four possible eigenvalues $e^{i m \pi/2}$ with $m=M$ mod $4$. In systems with an odd number of Cooper pairs (i.e., when $(N-1)/2$ is an odd integer), $p_x + i p_y$ and $p_x - i p_y$ ground states are degenerate in energy and have distinct $C_4$ eigenvalues $\pm i$ respectively. 

Since our NN variational optimization is entirely driven by energy minimization, the optimized NN wavefunction can be any superposition of (nearly) degenerate ground states, with distinct $C_4$ eigenvalues and opposite chiralities. In general, our NN wavefunction is a mixture of different $C_4$ symmetry eigenspaces, $\Psi=\sum_{m=0}^3\tilde\Psi_m$, where $\tilde\Psi_m$ picks up a phase $e^{i m\pi/2}$ under a $\pi/2$ rotation. To unmix different sectors, we project the NN wavefunction onto a given $C_4$ sector by making a proper superposition of $\Psi(\mathbf X)$ and its rotated copies, 
\begin{equation}
\label{eq:AM-projection}
    \tilde\Psi_m(\mathbf X)=\frac{1}{4}\sum_{k=0}^3 e^{-ikm\pi/2}\,\Psi(\mathbf R_{\pi/2}^k\mathbf X),
\end{equation}
where $\mathbf R_{\pi/2}$ denotes the rotation of all particle coordinates by the angle $\pi/2$, see App.~\ref{app:AM-projection} for a derivation.

\begin{figure}
    \centering
    \includegraphics[width=\linewidth]{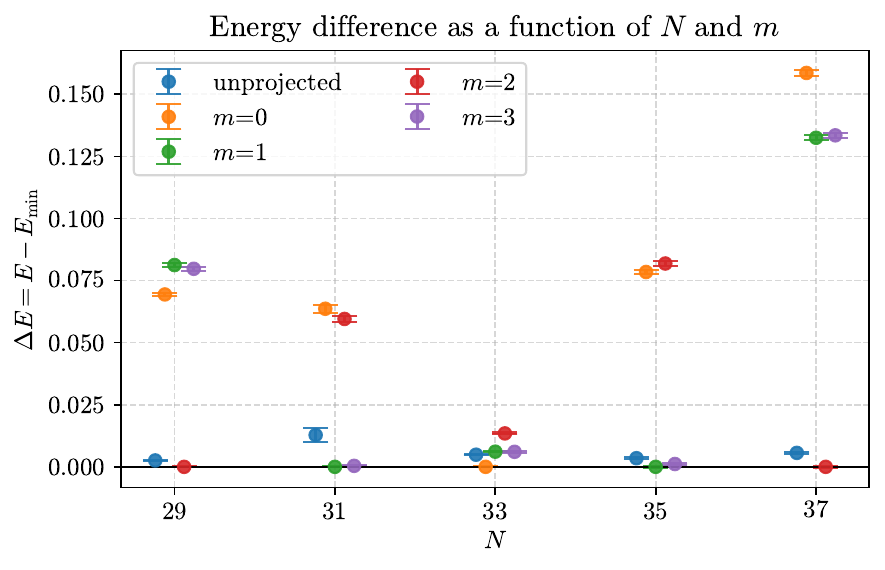}
    \caption{\justifying Angular-momentum sectors from $C_4$ projection (relative energies). For each particle number 
    $N$,
    the optimized wavefunction is projected onto the four $C_4$ eigenspaces $m=0,1,2,3$ and the \emph{relative} variational energy $\Delta E_m(N)\allowbreak\equiv\allowbreak E_m(N)\allowbreak-\min_{m'}\allowbreak E_{m'}(N)\allowbreak$ is shown. The lowest-energy sector follows the sequence $m=2,\allowbreak\ \{1,3\},\allowbreak\ 0,\allowbreak\ \{1,3\},\ 2$ (braces indicate degeneracy), consistent with chiral $p_x\pm ip_y$ pairing that shifts the total angular momentum by $\pm1$ upon adding a Cooper pair ($m\!\to\! m\pm1\ \mathrm{mod}\ 4$).}
    \label{fig:AM-pattern}
\end{figure}

In Fig.~\ref{fig:AM-pattern}, we compare the variational energies of the original NN wavefunction and its projections onto the $m=0,1,2,3$ angular-momentum channels. These projections improve the ground-state energy, and the angular momentum of the lowest-energy state follows the pattern $m\equiv \pm (N-1)/2\mod{4}$ for $N=29,31,33,35,37$, consistent with that of a chiral superconductor. Henceforth, we use the symmetrized wavefunction $\tilde\Psi_m$ with odd $(N-1)/2$, i.e. a $C_4$ eigenvalue of $e^{i m\pi/2}=\pm i$, as the variational ground state. The neural network spontaneously learns the two-fold degenerate ground-state within $\sim 3 \times 10^4$ training steps [see SM Fig.~\ref{fig:C4 proj training}]. The NQS has dominant, approximately equal weight in the angular momentum sectors corresponding to the chiral ground states [see SM Fig.~\ref{fig:C4-overlap}].


\begin{figure}
    \centering
    \begin{subfigure}[b]{0.23\textwidth}
         \centering
         \includegraphics[width=\textwidth]{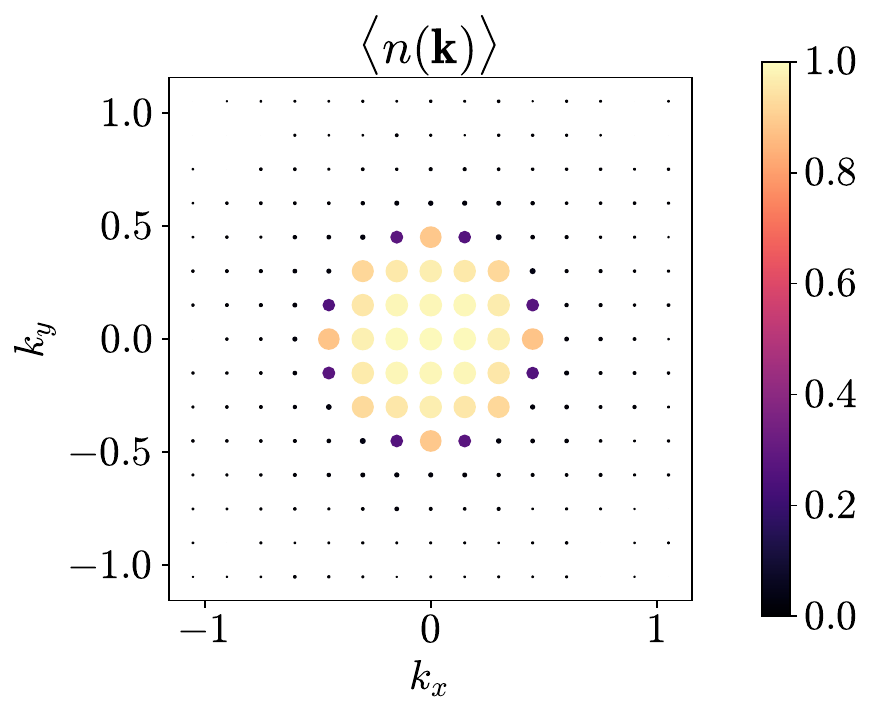}
     \end{subfigure}
     \hfill
     \begin{subfigure}[b]{0.23\textwidth}
         \centering
         \includegraphics[width=\textwidth]{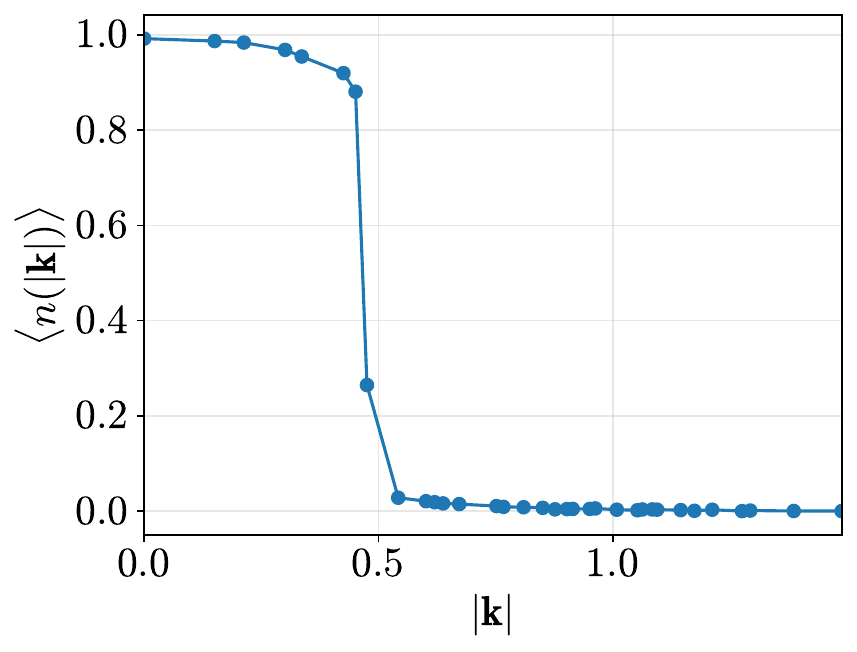}
     \end{subfigure}
    \caption{\justifying 
    Momentum distribution $\langle n(\mathbf k)\rangle$ for $N=31$ particles wavefunction. 
    Left: two-dimensional map of $n(\mathbf k)$, where both dot size and color indicate the occupation value. 
    Right: shell-averaged distribution $\langle n(|\mathbf k|)\rangle$. 
    The smeared occupation $\langle n(|\mathbf k|)\rangle$ around the Fermi surface is characteristic of pairing.
    }
    \label{fig:nk}
\end{figure}

\emph{Ground-state momentum distribution.}
The presence of pairing is also evidenced by the momentum distribution function $n(\mathbf k)=\langle \hat c^\dagger_{\mathbf k}\hat c_{\mathbf k}\rangle$. 
For the spinless non-interacting Fermi gas, all states inside the Fermi surface are strictly occupied with $n(\mathbf k)=1$, while those outside are empty with $n(\mathbf k)=0$. 
In contrast, in the superconducting ground state, the momentum distribution is broadened: instead of a sharp discontinuity, $n(\mathbf k)$ changes smoothly across the Fermi surface due to electron pairing.   
This behavior was indeed observed in $n(\mathbf k)$ calculated for the optimized NN state, see Fig.~\ref{fig:nk}.   
$n(|\mathbf k|)$ shows a smeared profile, consistent with BCS coherence factors 
of a fully gapped superconductor. Note that the occupation number of states considerably below the Fermi level is reduced from unity, consistent with the large binding energy.

{\em Off-diagonal long-range order---} 
To demonstrate the existence of superconductivity directly, we calculate the two-body reduced density matrix (2-RDM) of the many-body wavefunction. In real space the 2-RDM is
\begin{align}
    \rho^{(2)}(\mathbf x_1,\mathbf x_2;\mathbf x_1',\mathbf x_2')
    = \big\langle \hat c^\dagger_{\mathbf x_1}\hat c^\dagger_{\mathbf x_2}\hat c_{\mathbf x_2'}\hat c_{\mathbf x_1'}\big\rangle,
\end{align}
which satisfies $\mathrm{Tr}\,\rho^{(2)}=N(N\!-\!1)$. Equivalently, it is the partial trace of $|\Psi\rangle\!\langle\Psi|$ over $N\!-\!2$ particle coordinates: 
\begin{eqnarray}
\rho^{(2)} = N(N-1)\int d \tilde{\mathbf R}\Psi^*(\mathbf x_1,\mathbf x_2, \tilde{\mathbf R})\Psi(\mathbf x'_1,\mathbf x'_2, \tilde{\mathbf R}),
\end{eqnarray}
where, for brevity, we used the notation $\tilde{\mathbf R} \equiv (\mathbf{x}_3,\ldots,\mathbf{x}_N)$  to denote all other particle's coordinates.  

The defining feature of superconductivity due to electron pairing is that the 2-RDM $\rho^{(2)}(\mathbf x_1,\mathbf x_2;\mathbf x_1',\mathbf x_2')$ has a large eigenvalue $\lambda_0$ that is proportional to the particle number $N$ \cite{YangRMP1962}. This is the manifestation of macroscopic occupation of 
a Cooper paired state,
which is given by the corresponding eigenvector $\Phi_0(\mathbf x_1 , \mathbf x_2)$. For translationally invariant systems, $\Phi_0(\mathbf x_1 , \mathbf x_2)$ is a product of the center-of-mass part and the relative wavefunctions. 
In the case of zero center-of-mass momentum $\mathbf Q=0$ Cooper pairing as in our system,  $\Phi_0(\mathbf x_1 , \mathbf x_2) =\Phi_0(\mathbf x_1 - \mathbf x_2)$ reduces to a function of the relative coordinates only.

\begin{figure}[t]
    \centering
    \includegraphics[width=\linewidth]{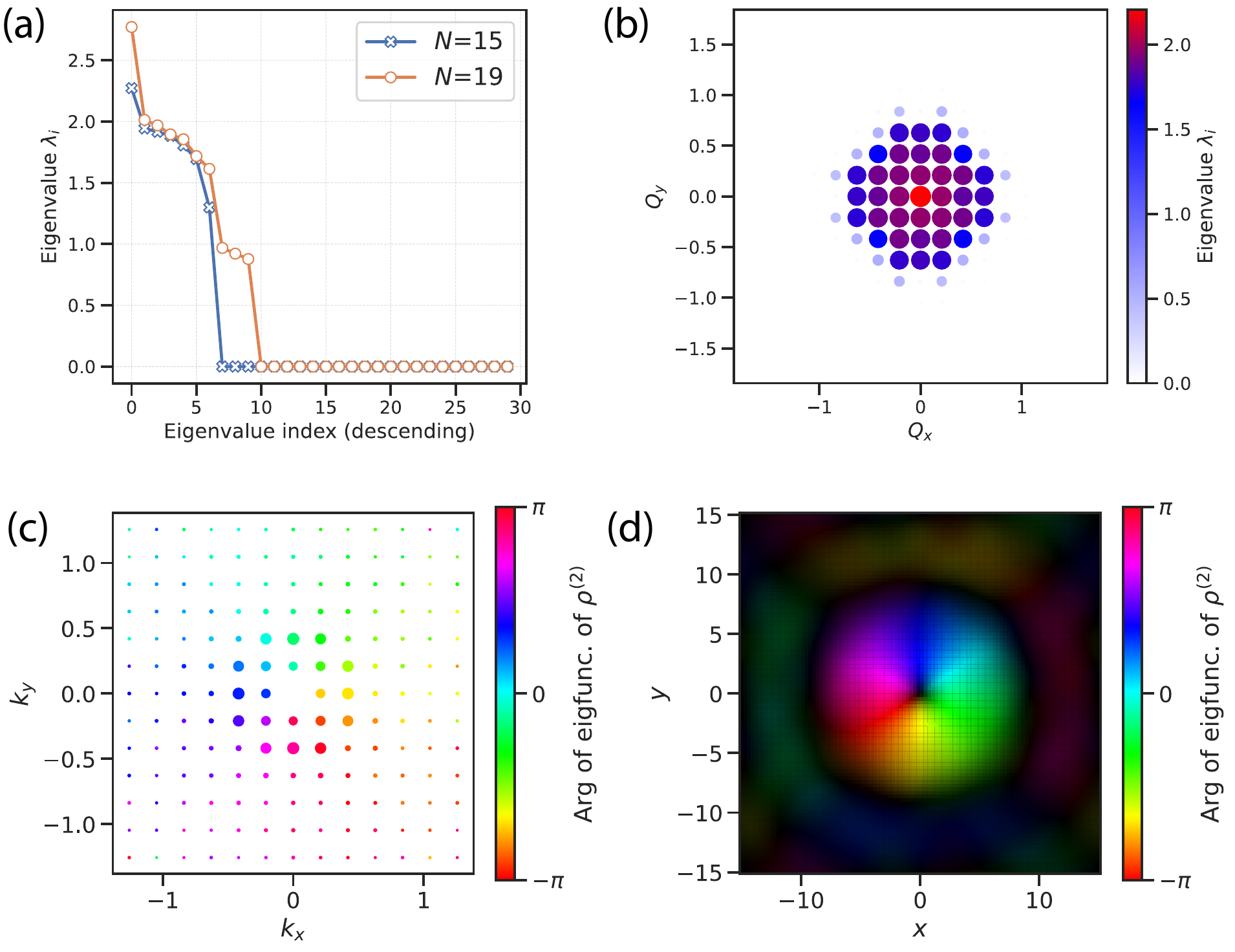}
    \caption{\justifying 2-RDM eigensystem at $U=-10$, $L=30$, $m=3$. 
    (a) Eigenspectrum of the 2-RDM for $N = 15$ and $19$ particles within the $\mathbf Q = 0$ sector.
    (b) Largest eigenvalue of the 2-RDM for $N = 15$ within each $\mathbf Q$ sector.
    (a) Leading eigenvector $\Phi_0(\mathbf k)$ for $N = 15$ on the discrete momentum grid; marker area $\propto |\Phi_0(\mathbf k)|$, color encodes Arg $\Phi_0(\mathbf k)$ (radians). The phase winds by $+2\pi$ around the origin, consistent with chiral $p_x\!+\! i p_y$ pairing.
    (b) Real-space pair wavefunction $\Phi_0(\mathbf x)=\allowbreak\sum_{\mathbf k}\allowbreak\Phi_0(\mathbf k)\allowbreak e^{i\mathbf k\cdot\mathbf x}$ for $N = 15$; intensity shows $|\Phi_0(\mathbf x)|$ and color its phase.}
    \label{fig:two-rdm}
\end{figure}

To extract the Cooper pair wavefunction $\Phi_0(\mathbf x)$, it is convenient to work with 2-RDM in $\mathbf k$ representation. Specifically, the 2-RDM in $\mathbf Q=0$ sector can be written in terms of the pair operator $\hat\Delta(\mathbf k)=\hat c_{-\mathbf k}\hat c_{\mathbf k}$:  
\begin{align}
    \Gamma_{\mathbf k,\mathbf k'}=\big\langle \hat\Delta^\dagger(\mathbf k)\,\hat\Delta(\mathbf k')\big\rangle .
\end{align}
Solving the eigenproblem
\begin{align}
    \sum_{\mathbf k'}\Gamma_{\mathbf k,\mathbf k'}\,\Phi_i(\mathbf k')=\lambda_i\,\Phi_i(\mathbf k)
\end{align}
yields an orthonormal set $\{\Phi_i\}$ with nonnegative eigenvalues $\{\lambda_i\}$. The leading eigenvalue $\lambda_0$ quantifies pair condensation (Penrose–Onsager criterion): $\lambda_0=\mathcal O(N)$ in a superconducting state, whereas all $\lambda_i=\mathcal O(1)$ in a normal state. 
Our numerical results [Fig.~\ref{fig:two-rdm}(a), (b)] display a dominant eigenvalue at $\mathbf Q = 0$ that grows with system size.
The leading eigenvector $\Phi_0(\mathbf k)$ corresponds to the Cooper pair wavefunction, i.e., the Fourier transform of $\Phi_0(\mathbf r)$ [{\it c.f.} Fig.~\ref{fig:two-rdm}(c) and (d)].

Our calculations of 2-RDM show that the largest eigenvalue is clearly detached from the remaining spectrum and continuum band. This separation provides direct evidence of ODLRO in the pair channel. As shown in Fig.~\ref{fig:two-rdm}, the corresponding eigenvector $\Phi_0(\mathbf k)$ exhibits a $2\pi$ phase winding around the origin, consistent with chiral $p_x+i p_y$ symmetry. To our knowledge, this is the first \emph{variational Monte Carlo} calculation that explicitly constructs and diagonalizes the two-body reduced density matrix to identify the Cooper pair wavefunction. Estimator details and numerical stabilization procedures are described in Appendix~\ref{app:observable-meas}.

To conclude, we have demonstrated that a self-attention NQS can faithfully represent the ground state of an interacting, time-reversal–breaking chiral $p_x+i p_y$ superconductor in a spin-polarized two-dimensional Fermi gas. 
Crucially, the variational ansatz enforces only fermionic antisymmetry, without imposing any Pfaffian, geminal, or other problem-specific pairing structure. 
All key features---including chiral $p_x+ip_y$ symmetry and its associated time-reversal symmetry breaking---are self-learned \emph{ab initio} by the NQS during variational optimization. 
To our knowledge, this is the first fully many-body, continuum-space simulation of chiral superconductivity. 
We present multiple lines of evidence: the even–odd effect in the pair-binding energy indicates that odd-$N$ states are energetically favored, and the sequence of ground-state total angular momentum across successive fillings—resolved via a Monte-Carlo–friendly $C_4$ symmetry projection—reflects the Cooper-pair chirality; moreover, the two-body RDM exhibits a single macroscopic eigenvalue at $\mathbf Q = 0$ with a leading eigenvector whose phase winds by $2\pi$ around the Fermi surface, directly revealing a chiral $p_x+i p_y$ pair wavefunction.

Our results show how transformer-based NQS can discover superconductivity from first principles. 
We demonstrated that {\it the attention-based NN is sufficiently expressive}, and emphasize that no new architectural innovations are required to find topological superconductivity. 
The key contribution lies in our deeper understanding of the expressivity of the NQS and our {\it development of the symmetry-projection method and eigendecomposition of the two-body RDM} to identify the chiral ground state.
This methodology lays a foundation for accurate, fully many-body descriptions of strongly correlated superconductivity emerging from electron–electron interactions, such as in graphene~\cite{Cao2018Apr,Chen2019Aug,Lu2019Oct,Arora2020Jul,Saito2020Sep,Park2021Feb,Hao2021Feb,Oh2021Dec,Zhou2021Oct,Kim2022Jun,Li2024Jul,han2025signatures}, transition-metal dichalcogenides~\cite{Xia2025Jan,Guo2025Jan}, and high-$T_{\rm c}$ cuprates~\cite{Bednorz1986highTc,Wu1987highTc}.


{\it Acknowledgments.---} It is our pleasure to thank Pierre-Antoine Graham and Daniele Guerci for informative discussions. This work made use of computing resources provided by subMIT at MIT Physics and by the National Science Foundation under Cooperative Agreement PHY-2019786. 
This work was primarily supported by the Air Force Office of Scientific Research under award number FA2386-24-1-4043. MG acknowledges support from the German Research Foundation under the Walter Benjamin program (Grant Agreement No. 526129603). CL, TO, HL and LF are grateful for the support from MISTI Global Seed Funds. LF was supported by a Simons Investigator Award from the Simons Foundation and the NSF through Award No. PHY-2425180.

\bibliography{reference.bib}

@article{carleo2017solving,
  title={Solving the quantum many-body problem with artificial neural networks},
  author={Carleo, Giuseppe and Troyer, Matthias},
  journal={Science},
  volume={355},
  number={6325},
  pages={602--606},
  year={2017},
  publisher={American Association for the Advancement of Science}
}

@article{pfau2020ab,
  title={Ab initio solution of the many-electron Schr{\"o}dinger equation with deep neural networks},
  author={Pfau, David and Spencer, James S and Matthews, Alexander GDG and Foulkes, W Matthew C},
  journal={Physical review research},
  volume={2},
  number={3},
  pages={033429},
  year={2020},
  publisher={APS}
}

@article{hermann2020deep,
  title={Deep-neural-network solution of the electronic Schr{\"o}dinger equation},
  author={Hermann, Jan and Sch{\"a}tzle, Zeno and No{\'e}, Frank},
  journal={Nature Chemistry},
  volume={12},
  number={10},
  pages={891--897},
  year={2020},
  publisher={Nature Publishing Group UK London}
}

@article{hohenberg1964inhomogeneous,
  title={Inhomogeneous electron gas},
  author={Hohenberg, Pierre and Kohn, Walter},
  journal={Physical review},
  volume={136},
  number={3B},
  pages={B864},
  year={1964},
  publisher={APS}
}

@article{kohn1965self,
  title={Self-consistent equations including exchange and correlation effects},
  author={Kohn, Walter and Sham, Lu Jeu},
  journal={Physical review},
  volume={140},
  number={4A},
  pages={A1133},
  year={1965},
  publisher={APS}
}

@article{ceperley1980ground,
  title={Ground state of the electron gas by a stochastic method},
  author={Ceperley, David M and Alder, Berni J},
  journal={Physical review letters},
  volume={45},
  number={7},
  pages={566},
  year={1980},
  publisher={APS}
}

@article{foulkes2001quantum,
  title={Quantum Monte Carlo simulations of solids},
  author={Foulkes, William MC and Mitas, Lubos and Needs, RJ and Rajagopal, Guna},
  journal={Reviews of Modern Physics},
  volume={73},
  number={1},
  pages={33},
  year={2001},
  publisher={APS}
}

@book{becca2017quantum,
  title={Quantum Monte Carlo approaches for correlated systems},
  author={Becca, Federico and Sorella, Sandro},
  year={2017},
  publisher={Cambridge University Press}
}

@article{white1992density,
  title={Density matrix formulation for quantum renormalization groups},
  author={White, Steven R},
  journal={Physical review letters},
  volume={69},
  number={19},
  pages={2863},
  year={1992},
  publisher={APS}
}

@article{verstraete2023density,
  title={Density matrix renormalization group, 30 years on},
  author={Verstraete, Frank and Nishino, Tomotoshi and Schollw{\"o}ck, Ulrich and Ba{\~n}uls, Mari Carmen and Chan, Garnet K and Stoudenmire, Miles E},
  journal={Nature Reviews Physics},
  volume={5},
  number={5},
  pages={273--276},
  year={2023},
  publisher={Nature Publishing Group UK London}
}

@article{kim2024neural,
  title={Neural-network quantum states for ultra-cold Fermi gases},
  author={Kim, Jane and Pescia, Gabriel and Fore, Bryce and Nys, Jannes and Carleo, Giuseppe and Gandolfi, Stefano and Hjorth-Jensen, Morten and Lovato, Alessandro},
  journal={Communications Physics},
  volume={7},
  number={1},
  pages={148},
  year={2024},
  publisher={Nature Publishing Group UK London}
}

@article{lou2024neural,
  title={Neural wave functions for superfluids},
  author={Lou, Wan Tong and Sutterud, Halvard and Cassella, Gino and Foulkes, W Matthew C and Knolle, Johannes and Pfau, David and Spencer, James S},
  journal={Physical Review X},
  volume={14},
  number={2},
  pages={021030},
  year={2024},
  publisher={APS}
}

@article{sorella1998green,
  title={Green function Monte Carlo with stochastic reconfiguration},
  author={Sorella, Sandro},
  journal={Physical review letters},
  volume={80},
  number={20},
  pages={4558},
  year={1998},
  publisher={APS}
}

@article{stokes2020quantum,
  title={Quantum natural gradient},
  author={Stokes, James and Izaac, Josh and Killoran, Nathan and Carleo, Giuseppe},
  journal={Quantum},
  volume={4},
  pages={269},
  year={2020},
  publisher={Verein zur F{\"o}rderung des Open Access Publizierens in den Quantenwissenschaften}
}

@article{amari1998natural,
  title={Natural gradient works efficiently in learning},
  author={Amari, Shun-Ichi},
  journal={Neural computation},
  volume={10},
  number={2},
  pages={251--276},
  year={1998},
  publisher={MIT Press}
}

@inproceedings{martens2015optimizing,
  title={Optimizing neural networks with kronecker-factored approximate curvature},
  author={Martens, James and Grosse, Roger},
  booktitle={International conference on machine learning},
  pages={2408--2417},
  year={2015},
  organization={PMLR}
}

@article{vaswani2017attention,
  title={Attention is all you need},
  author={Vaswani, Ashish and Shazeer, Noam and Parmar, Niki and Uszkoreit, Jakob and Jones, Llion and Gomez, Aidan N and Kaiser, {\L}ukasz and Polosukhin, Illia},
  journal={Advances in neural information processing systems},
  volume={30},
  year={2017}
}

@article{geier2025attention,
  title = {Self-attention neural network for solving correlated electron problems in solids},
  author = {Geier, Max and Nazaryan, Khachatur and Zaklama, Timothy and Fu, Liang},
  journal = {Phys. Rev. B},
  volume = {112},
  issue = {4},
  pages = {045119},
  numpages = {16},
  year = {2025},
  month = {Jul},
  publisher = {American Physical Society},
  doi = {10.1103/qxc3-bkc7},
  url = {https://link.aps.org/doi/10.1103/qxc3-bkc7}
}

@article{han2025signatures,
	author = {Han, Tonghang and Lu, Zhengguang and Hadjri, Zach and Shi, Lihan and Wu, Zhenghan and Xu, Wei and Yao, Yuxuan and Cotten, Armel A. and Sharifi Sedeh, Omid and Weldeyesus, Henok and Yang, Jixiang and Seo, Junseok and Ye, Shenyong and Zhou, Muyang and Liu, Haoyang and Shi, Gang and Hua, Zhenqi and Watanabe, Kenji and Taniguchi, Takashi and Xiong, Peng and Zumb{\"u}hl, Dominik M. and Fu, Liang and Ju, Long},
	date = {2025/07/01},
	doi = {10.1038/s41586-025-09169-7},
	id = {Han2025},
	isbn = {1476-4687},
	journal = {Nature},
	number = {8072},
	pages = {654--661},
	title = {Signatures of chiral superconductivity in rhombohedral graphene},
	url = {https://doi.org/10.1038/s41586-025-09169-7},
	volume = {643},
	year = {2025}}

@article{Cao2018Apr,
	author = {Cao, Yuan and Fatemi, Valla and Fang, Shiang and Watanabe, Kenji and Taniguchi, Takashi and Kaxiras, Efthimios and Jarillo-Herrero, Pablo},
	title = {{Unconventional superconductivity in magic-angle graphene superlattices}},
	journal = {Nature},
	volume = {556},
	pages = {43--50},
	year = {2018},
	month = apr,
	issn = {1476-4687},
	publisher = {Nature Publishing Group},
	doi = {10.1038/nature26160}
}

@article{Oh2021Dec,
	author = {Oh, Myungchul and Nuckolls, Kevin P. and Wong, Dillon and Lee, Ryan L. and Liu, Xiaomeng and Watanabe, Kenji and Taniguchi, Takashi and Yazdani, Ali},
	title = {{Evidence for unconventional superconductivity in twisted bilayer graphene}},
	journal = {Nature},
	volume = {600},
	pages = {240--245},
	year = {2021},
	month = dec,
	issn = {1476-4687},
	publisher = {Nature Publishing Group},
	doi = {10.1038/s41586-021-04121-x}
}

@article{Lu2019Oct,
	author = {Lu, Xiaobo and Stepanov, Petr and Yang, Wei and Xie, Ming and Aamir, Mohammed Ali and Das, Ipsita and Urgell, Carles and Watanabe, Kenji and Taniguchi, Takashi and Zhang, Guangyu and Bachtold, Adrian and MacDonald, Allan H. and Efetov, Dmitri K.},
	title = {{Superconductors, orbital magnets and correlated states in magic-angle bilayer graphene}},
	journal = {Nature},
	volume = {574},
	pages = {653--657},
	year = {2019},
	month = oct,
	issn = {1476-4687},
	publisher = {Nature Publishing Group},
	doi = {10.1038/s41586-019-1695-0}
}

@article{Saito2020Sep,
	author = {Saito, Yu and Ge, Jingyuan and Watanabe, Kenji and Taniguchi, Takashi and Young, Andrea F.},
	title = {{Independent superconductors and correlated insulators in twisted bilayer graphene}},
	journal = {Nat. Phys.},
	volume = {16},
	pages = {926--930},
	year = {2020},
	month = sep,
	issn = {1745-2481},
	publisher = {Nature Publishing Group},
	doi = {10.1038/s41567-020-0928-3}
}

@article{Li2024Jul,
	author = {Li, Chushan and Xu, Fan and Li, Bohao and Li, Jiayi and Li, Guoan and Watanabe, Kenji and Taniguchi, Takashi and Tong, Bingbing and Shen, Jie and Lu, Li and Jia, Jinfeng and Wu, Fengcheng and Liu, Xiaoxue and Li, Tingxin},
	title = {{Tunable superconductivity in electron- and hole-doped Bernal bilayer graphene}},
	journal = {Nature},
	volume = {631},
	pages = {300--306},
	year = {2024},
	month = jul,
	issn = {1476-4687},
	publisher = {Nature Publishing Group},
	doi = {10.1038/s41586-024-07584-w}
}

@article{Zhou2021Oct,
	author = {Zhou, Haoxin and Xie, Tian and Taniguchi, Takashi and Watanabe, Kenji and Young, Andrea F.},
	title = {{Superconductivity in rhombohedral trilayer graphene}},
	journal = {Nature},
	volume = {598},
	pages = {434--438},
	year = {2021},
	month = oct,
	issn = {1476-4687},
	publisher = {Nature Publishing Group},
	doi = {10.1038/s41586-021-03926-0}
}

@article{Kim2022Jun,
	author = {Kim, Hyunjin and Choi, Youngjoon and Lewandowski, Cyprian and Thomson, Alex and Zhang, Yiran and Polski, Robert and Watanabe, Kenji and Taniguchi, Takashi and Alicea, Jason and Nadj-Perge, Stevan},
	title = {{Evidence for unconventional superconductivity in twisted trilayer graphene}},
	journal = {Nature},
	volume = {606},
	pages = {494--500},
	year = {2022},
	month = jun,
	issn = {1476-4687},
	publisher = {Nature Publishing Group},
	doi = {10.1038/s41586-022-04715-z}
}

@article{Park2021Feb,
	author = {Park, Jeong Min and Cao, Yuan and Watanabe, Kenji and Taniguchi, Takashi and Jarillo-Herrero, Pablo},
	title = {{Tunable strongly coupled superconductivity in magic-angle twisted trilayer graphene}},
	journal = {Nature},
	volume = {590},
	pages = {249--255},
	year = {2021},
	month = feb,
	issn = {1476-4687},
	publisher = {Nature Publishing Group},
	doi = {10.1038/s41586-021-03192-0}
}

@article{Arora2020Jul,
	author = {Arora, Harpreet Singh and Polski, Robert and Zhang, Yiran and Thomson, Alex and Choi, Youngjoon and Kim, Hyunjin and Lin, Zhong and Wilson, Ilham Zaky and Xu, Xiaodong and Chu, Jiun-Haw and Watanabe, Kenji and Taniguchi, Takashi and Alicea, Jason and Nadj-Perge, Stevan},
	title = {{Superconductivity in metallic twisted bilayer graphene stabilized by WSe2}},
	journal = {Nature},
	volume = {583},
	pages = {379--384},
	year = {2020},
	month = jul,
	issn = {1476-4687},
	publisher = {Nature Publishing Group},
	doi = {10.1038/s41586-020-2473-8}
}

@article{Hao2021Feb,
	author = {Hao, Zeyu and Zimmerman, A. M. and Ledwith, Patrick and Khalaf, Eslam and Najafabadi, Danial Haie and Watanabe, Kenji and Taniguchi, Takashi and Vishwanath, Ashvin and Kim, Philip},
	title = {{Electric field{\textendash}tunable superconductivity in alternating-twist magic-angle trilayer graphene}},
	journal = {Science},
	volume = {371},
	number = {6534},
	pages = {1133--1138},
	year = {2021},
	month = feb,
	issn = {0036-8075},
	publisher = {American Association for the Advancement of Science},
	doi = {10.1126/science.abg0399}
}

@article{Chen2019Aug,
	author = {Chen, Guorui and Sharpe, Aaron L. and Gallagher, Patrick and Rosen, Ilan T. and Fox, Eli J. and Jiang, Lili and Lyu, Bosai and Li, Hongyuan and Watanabe, Kenji and Taniguchi, Takashi and Jung, Jeil and Shi, Zhiwen and Goldhaber-Gordon, David and Zhang, Yuanbo and Wang, Feng},
	title = {{Signatures of tunable superconductivity in a trilayer graphene moir{\ifmmode\acute{e}\else\'{e}\fi} superlattice}},
	journal = {Nature},
	volume = {572},
	pages = {215--219},
	year = {2019},
	month = aug,
	issn = {1476-4687},
	publisher = {Nature Publishing Group},
	doi = {10.1038/s41586-019-1393-y}
}

@article{Xia2025Jan,
	author = {Xia, Yiyu and Han, Zhongdong and Watanabe, Kenji and Taniguchi, Takashi and Shan, Jie and Mak, Kin Fai},
	date = {2025/01/01},
	doi = {10.1038/s41586-024-08116-2},
	id = {Xia2025},
	isbn = {1476-4687},
	journal = {Nature},
	number = {8047},
	pages = {833--838},
	title = {{Superconductivity in twisted bilayer WSe2}},
	url = {https://doi.org/10.1038/s41586-024-08116-2},
	volume = {637},
	year = {2025}}

@article{Wu1987highTc,
  title = {{Superconductivity at 93 K in a new mixed-phase Y-Ba-Cu-O compound system at ambient pressure}},
  author = {Wu, M. K. and Ashburn, J. R. and Torng, C. J. and Hor, P. H. and Meng, R. L. and Gao, L. and Huang, Z. J. and Wang, Y. Q. and Chu, C. W.},
  journal = {Phys. Rev. Lett.},
  volume = {58},
  issue = {9},
  pages = {908--910},
  numpages = {0},
  year = {1987},
  month = {Mar},
  publisher = {American Physical Society},
  doi = {10.1103/PhysRevLett.58.908},
  url = {https://link.aps.org/doi/10.1103/PhysRevLett.58.908}
}

@article{Bednorz1986highTc,
	author = {Bednorz, J. G. and M{\"u}ller, K. A.},
	date = {1986/06/01},
	doi = {10.1007/BF01303701},
	id = {Bednorz1986},
	isbn = {1431-584X},
	journal = {Zeitschrift f\"ur Physik B Condensed Matter},
	number = {2},
	pages = {189--193},
	title = {{Possible high Tc superconductivity in the Ba-La-Cu-O system}},
	url = {https://doi.org/10.1007/BF01303701},
	volume = {64},
	year = {1986}}

@article{von2022self,
  title={A self-attention ansatz for ab-initio quantum chemistry},
  author={von Glehn, Ingrid and Spencer, James S and Pfau, David},
  journal={arXiv preprint arXiv:2211.13672},
  year={2022}
}

@article{teng2025solving,
  title={Solving the fractional quantum Hall problem with self-attention neural network},
  author={Teng, Yi and Dai, David D and Fu, Liang},
  journal={Physical Review B},
  volume={111},
  number={20},
  pages={205117},
  year={2025},
  publisher={APS}
}

@article{li2022ab,
  title={Ab initio calculation of real solids via neural network ansatz},
  author={Li, Xiang and Li, Zhe and Chen, Ji},
  journal={Nature Communications},
  volume={13},
  number={1},
  pages={7895},
  year={2022},
  publisher={Nature Publishing Group UK London}
}

@article{scherbela2024towards,
  title={Towards a transferable fermionic neural wavefunction for molecules},
  author={Scherbela, Michael and Gerard, Leon and Grohs, Philipp},
  journal={Nature Communications},
  volume={15},
  number={1},
  pages={120},
  year={2024},
  publisher={Nature Publishing Group UK London}
}

@article{rende2024fine,
  title={Fine-tuning neural network quantum states},
  author={Rende, Riccardo and Goldt, Sebastian and Becca, Federico and Viteritti, Luciano Loris},
  journal={Physical Review Research},
  volume={6},
  number={4},
  pages={043280},
  year={2024},
  publisher={APS}
}

@article{cybenko1989approximation,
  title={Approximation by superpositions of a sigmoidal function},
  author={Cybenko, George},
  journal={Mathematics of control, signals and systems},
  volume={2},
  number={4},
  pages={303--314},
  year={1989},
  publisher={Springer}
}

@article{funahashi1989approximate,
  title={On the approximate realization of continuous mappings by neural networks},
  author={Funahashi, Ken-Ichi},
  journal={Neural networks},
  volume={2},
  number={3},
  pages={183--192},
  year={1989},
  publisher={Elsevier}
}

@article{hornik1989multilayer,
  title={Multilayer feedforward networks are universal approximators},
  author={Hornik, Kurt and Stinchcombe, Maxwell and White, Halbert},
  journal={Neural networks},
  volume={2},
  number={5},
  pages={359--366},
  year={1989},
  publisher={Elsevier}
}

@article{read2000paired,
  title={Paired states of fermions in two dimensions with breaking of parity and time-reversal symmetries and the fractional quantum Hall effect},
  author={Read, Nicholas and Green, Dmitry},
  journal={Physical Review B},
  volume={61},
  number={15},
  pages={10267},
  year={2000},
  publisher={APS}
}

@article{fu2010electron,
  title={Electron teleportation via Majorana bound states in a mesoscopic superconductor},
  author={Fu, Liang},
  journal={Physical review letters},
  volume={104},
  number={5},
  pages={056402},
  year={2010},
  publisher={APS}
}

@book{schrieffer2018theory,
  title={Theory of superconductivity},
  author={Schrieffer, J Robert},
  year={2018},
  publisher={CRC press}
}

@article{lu2010superconducting,
  title={Superconducting order parameter for the even-denominator fractional quantum Hall effect},
  author={Lu, Hantao and Das Sarma, S and Park, Kwon},
  journal={Physical Review B—Condensed Matter and Materials Physics},
  volume={82},
  number={20},
  pages={201303},
  year={2010},
  publisher={APS}
}

@article{Pescia2024MessagePassing,
  title = {Message-passing neural quantum states for the homogeneous electron gas},
  author = {Pescia, Gabriel and Nys, Jannes and Kim, Jane and Lovato, Alessandro and Carleo, Giuseppe},
  journal = {Phys. Rev. B},
  volume = {110},
  issue = {3},
  pages = {035108},
  numpages = {11},
  year = {2024},
  month = {Jul},
  publisher = {American Physical Society},
  doi = {10.1103/PhysRevB.110.035108},
  url = {https://link.aps.org/doi/10.1103/PhysRevB.110.035108}
}

@misc{luo2024simulatingmoire,
      title={Simulating moir\'e quantum matter with neural network}, 
      author={Di Luo and David D. Dai and Liang Fu},
      year={2024},
      eprint={2406.17645},
      archivePrefix={arXiv},
      primaryClass={cond-mat.str-el},
      url={https://arxiv.org/abs/2406.17645}, 
}

@article{XLi2025Moire,
	author = {Li, Xiang and Qian, Yubing and Ren, Weiluo and Xu, Yang and Chen, Ji},
	date = {2025/09/02},
	doi = {10.1038/s42005-025-02282-z},
	id = {Li2025},
	isbn = {2399-3650},
	journal = {Communications Physics},
	number = {1},
	pages = {364},
	title = {Emergent Wigner phases in moir{\'e}superlattice from deep learning},
	url = {https://doi.org/10.1038/s42005-025-02282-z},
	volume = {8},
	year = {2025}}

@article{Smith2024Unified,
  title = {Unified Variational Approach Description of Ground-State Phases of the Two-Dimensional Electron Gas},
  author = {Smith, Conor and Chen, Yixiao and Levy, Ryan and Yang, Yubo and Morales, Miguel A. and Zhang, Shiwei},
  journal = {Phys. Rev. Lett.},
  volume = {133},
  issue = {26},
  pages = {266504},
  numpages = {6},
  year = {2024},
  month = {Dec},
  publisher = {American Physical Society},
  doi = {10.1103/PhysRevLett.133.266504},
  url = {https://link.aps.org/doi/10.1103/PhysRevLett.133.266504}
}

@article{Kim2024UltraCold,
	author = {Kim, Jane and Pescia, Gabriel and Fore, Bryce and Nys, Jannes and Carleo, Giuseppe and Gandolfi, Stefano and Hjorth-Jensen, Morten and Lovato, Alessandro},
	date = {2024/05/08},
	doi = {10.1038/s42005-024-01613-w},
	id = {Kim2024},
	isbn = {2399-3650},
	journal = {Communications Physics},
	number = {1},
	pages = {148},
	title = {Neural-network quantum states for ultra-cold Fermi gases},
	url = {https://doi.org/10.1038/s42005-024-01613-w},
	volume = {7},
	year = {2024}}

@misc{luo2023pairing,
      title={Pairing-based graph neural network for simulating quantum materials}, 
      author={Di Luo and David D. Dai and Liang Fu},
      year={2023},
      eprint={2311.02143},
      archivePrefix={arXiv},
      primaryClass={cond-mat.str-el},
      url={https://arxiv.org/abs/2311.02143}, 
}

@article{LuoPRL2019Backflow,
  title = {Backflow Transformations via Neural Networks for Quantum Many-Body Wave Functions},
  author = {Luo, Di and Clark, Bryan K.},
  journal = {Phys. Rev. Lett.},
  volume = {122},
  issue = {22},
  pages = {226401},
  numpages = {6},
  year = {2019},
  month = {Jun},
  publisher = {American Physical Society},
  doi = {10.1103/PhysRevLett.122.226401},
  url = {https://link.aps.org/doi/10.1103/PhysRevLett.122.226401}
}

@article{KwonPRB1993,
  title = {Effects of three-body and backflow correlations in the two-dimensional electron gas},
  author = {Kwon, Yongkyung and Ceperley, D. M. and Martin, Richard M.},
  journal = {Phys. Rev. B},
  volume = {48},
  issue = {16},
  pages = {12037--12046},
  numpages = {0},
  year = {1993},
  month = {Oct},
  publisher = {American Physical Society},
  doi = {10.1103/PhysRevB.48.12037},
  url = {https://link.aps.org/doi/10.1103/PhysRevB.48.12037}
}

@article{Cohen1956Backflow,
  title = {Energy Spectrum of the Excitations in Liquid Helium},
  author = {Feynman, R. P. and Cohen, Michael},
  journal = {Phys. Rev.},
  volume = {102},
  issue = {5},
  pages = {1189--1204},
  numpages = {0},
  year = {1956},
  month = {Jun},
  publisher = {American Physical Society},
  doi = {10.1103/PhysRev.102.1189},
  url = {https://link.aps.org/doi/10.1103/PhysRev.102.1189}
}

@inproceedings{Gao2023Jul,
author = {Gao, Nicholas and G\"{u}nnemann, Stephan},
title = {Generalizing neural wave functions},
year = {2023},
publisher = {JMLR.org},
booktitle = {Proceedings of the 40th International Conference on Machine Learning},
articleno = {432},
numpages = {19},
location = {Honolulu, Hawaii, USA},
series = {ICML'23}
}

@misc{gerard2024transferableSolids,
      title={Transferable Neural Wavefunctions for Solids}, 
      author={Leon Gerard and Michael Scherbela and Halvard Sutterud and Matthew Foulkes and Philipp Grohs},
      year={2024},
      eprint={2405.07599},
      archivePrefix={arXiv},
      primaryClass={physics.comp-ph},
      url={https://arxiv.org/abs/2405.07599}, 
}

@misc{gu2025solvinghubbardmodelneural,
      title={Solving the Hubbard model with Neural Quantum States}, 
      author={Yuntian Gu and Wenrui Li and Heng Lin and Bo Zhan and Ruichen Li and Yifei Huang and Di He and Yantao Wu and Tao Xiang and Mingpu Qin and Liwei Wang and Dingshun Lv},
      year={2025},
      eprint={2507.02644},
      archivePrefix={arXiv},
      primaryClass={cond-mat.str-el},
      url={https://arxiv.org/abs/2507.02644}, 
}

@article{viteritti2023transformer,
  title = {Transformer Variational Wave Functions for Frustrated Quantum Spin Systems},
  author = {Viteritti, Luciano Loris and Rende, Riccardo and Becca, Federico},
  journal = {Phys. Rev. Lett.},
  volume = {130},
  issue = {23},
  pages = {236401},
  numpages = {6},
  year = {2023},
  month = {Jun},
  publisher = {American Physical Society},
  doi = {10.1103/PhysRevLett.130.236401},
  url = {https://link.aps.org/doi/10.1103/PhysRevLett.130.236401}
}

@article{Wilson2023Jun,
	author = {Wilson, Max and Moroni, Saverio and Holzmann, Markus and Gao, Nicholas and Wudarski, Filip and Vegge, Tejs and Bhowmik, Arghya},
	title = {{Neural network ansatz for periodic wave functions and the homogeneous electron gas}},
	journal = {Phys. Rev. B},
	volume = {107},
	number = {23},
	pages = {235139},
	year = {2023},
	month = jun,
	publisher = {American Physical Society},
	doi = {10.1103/PhysRevB.107.235139}
}

@article{Cassella2023Jan,
	author = {Cassella, Gino and Sutterud, Halvard and Azadi, Sam and Drummond, N. D. and Pfau, David and Spencer, James S. and Foulkes, W. M. C.},
	title = {{Discovering Quantum Phase Transitions with Fermionic Neural Networks}},
	journal = {Phys. Rev. Lett.},
	volume = {130},
	number = {3},
	pages = {036401},
	year = {2023},
	month = jan,
	publisher = {American Physical Society},
	doi = {10.1103/PhysRevLett.130.036401}
}

@misc{foster2025abinitiofoundationmodel,
      title={An ab initio foundation model of wavefunctions that accurately describes chemical bond breaking}, 
      author={Adam Foster and Zeno Schätzle and P. Bernát Szabó and Lixue Cheng and Jonas Köhler and Gino Cassella and Nicholas Gao and Jiawei Li and Frank Noé and Jan Hermann},
      year={2025},
      eprint={2506.19960},
      archivePrefix={arXiv},
      primaryClass={physics.chem-ph},
      url={https://arxiv.org/abs/2506.19960}, 
}

@article{Li2024ForwardLaplacian,
	author = {Li, Ruichen and Ye, Haotian and Jiang, Du and Wen, Xuelan and Wang, Chuwei and Li, Zhe and Li, Xiang and He, Di and Chen, Ji and Ren, Weiluo and Wang, Liwei},
	date = {2024/02/01},
	doi = {10.1038/s42256-024-00794-x},
	id = {Li2024},
	isbn = {2522-5839},
	journal = {Nature Machine Intelligence},
	number = {2},
	pages = {209--219},
	title = {A computational framework for neural network-based variational Monte Carlo with Forward Laplacian},
	url = {https://doi.org/10.1038/s42256-024-00794-x},
	volume = {6},
	year = {2024}}

@article{Li2022Solids,
	author = {Li, Xiang and Li, Zhe and Chen, Ji},
	date = {2022/12/22},
	doi = {10.1038/s41467-022-35627-1},
	id = {Li2022},
	isbn = {2041-1723},
	journal = {Nature Communications},
	number = {1},
	pages = {7895},
	title = {Ab initio calculation of real solids via neural network ansatz},
	url = {https://doi.org/10.1038/s41467-022-35627-1},
	volume = {13},
	year = {2022}}

@article{Hermann2019PauliNet,
	author = {Hermann, Jan and Sch{\"a}tzle, Zeno and No{\'e}, Frank},
	date = {2020/10/01},
	doi = {10.1038/s41557-020-0544-y},
	id = {Hermann2020},
	isbn = {1755-4349},
	journal = {Nature Chemistry},
	number = {10},
	pages = {891--897},
	title = {Deep-neural-network solution of the electronic Schr{\"o}dinger equation},
	url = {https://doi.org/10.1038/s41557-020-0544-y},
	volume = {12},
	year = {2020}}

@article{Hermann2023Review,
	author = {Hermann, Jan and Spencer, James and Choo, Kenny and Mezzacapo, Antonio and Foulkes, W. M. C. and Pfau, David and Carleo, Giuseppe and No{\'e}, Frank},
	date = {2023/10/01},
	doi = {10.1038/s41570-023-00516-8},
	id = {Hermann2023},
	isbn = {2397-3358},
	journal = {Nature Reviews Chemistry},
	number = {10},
	pages = {692--709},
	title = {Ab initio quantum chemistry with neural-network wavefunctions},
	url = {https://doi.org/10.1038/s41570-023-00516-8},
	volume = {7},
	year = {2023}}

@software{periodicwave_github,
  author = {Max Geier, Khachatur Nazaryan},
  title = {{PeriodicWave}},
  url = {http://github.com/mg607/periodicwave},
  year = {2025},
}

@article{Han2025Chiral,
	author = {Han, Tonghang and Lu, Zhengguang and Hadjri, Zach and Shi, Lihan and Wu, Zhenghan and Xu, Wei and Yao, Yuxuan and Cotten, Armel A. and Sharifi Sedeh, Omid and Weldeyesus, Henok and Yang, Jixiang and Seo, Junseok and Ye, Shenyong and Zhou, Muyang and Liu, Haoyang and Shi, Gang and Hua, Zhenqi and Watanabe, Kenji and Taniguchi, Takashi and Xiong, Peng and Zumb{\"u}hl, Dominik M. and Fu, Liang and Ju, Long},
	date = {2025/07/01},
	doi = {10.1038/s41586-025-09169-7},
	id = {Han2025},
	isbn = {1476-4687},
	journal = {Nature},
	number = {8072},
	pages = {654--661},
	title = {Signatures of chiral superconductivity in rhombohedral graphene},
	url = {https://doi.org/10.1038/s41586-025-09169-7},
	volume = {643},
	year = {2025}}

@article{SchollwockRMP2005,
  title = {The density-matrix renormalization group},
  author = {Schollw\"ock, U.},
  journal = {Rev. Mod. Phys.},
  volume = {77},
  issue = {1},
  pages = {259--315},
  numpages = {0},
  year = {2005},
  month = {Apr},
  publisher = {American Physical Society},
  doi = {10.1103/RevModPhys.77.259},
  url = {https://link.aps.org/doi/10.1103/RevModPhys.77.259}
}

@article{YangRMP1962,
  title = {Concept of Off-Diagonal Long-Range Order and the Quantum Phases of Liquid He and of Superconductors},
  author = {Yang, C. N.},
  journal = {Rev. Mod. Phys.},
  volume = {34},
  issue = {4},
  pages = {694--704},
  numpages = {0},
  year = {1962},
  month = {Oct},
  publisher = {American Physical Society},
  doi = {10.1103/RevModPhys.34.694},
  url = {https://link.aps.org/doi/10.1103/RevModPhys.34.694}
}

@article{cao2018insulator,
	author = {Cao, Yuan and Fatemi, Valla and Demir, Ahmet and Fang, Shiang and Tomarken, Spencer L. and Luo, Jason Y. and Sanchez-Yamagishi, Javier D. and Watanabe, Kenji and Taniguchi, Takashi and Kaxiras, Efthimios and Ashoori, Ray C. and Jarillo-Herrero, Pablo},
	date = {2018/04/01},
	doi = {10.1038/nature26154},
	id = {Cao2018},
	isbn = {1476-4687},
	journal = {Nature},
	number = {7699},
	pages = {80--84},
	title = {Correlated insulator behaviour at half-filling in magic-angle graphene superlattices},
	url = {https://doi.org/10.1038/nature26154},
	volume = {556},
	year = {2018}}

@article{Yankowitz2019superconductivity,
author = {Matthew Yankowitz  and Shaowen Chen  and Hryhoriy Polshyn  and Yuxuan Zhang  and K. Watanabe  and T. Taniguchi  and David Graf  and Andrea F. Young  and Cory R. Dean },
title = {Tuning superconductivity in twisted bilayer graphene},
journal = {Science},
volume = {363},
number = {6431},
pages = {1059-1064},
year = {2019},
doi = {10.1126/science.aav1910},
URL = {https://www.science.org/doi/abs/10.1126/science.aav1910},
eprint = {https://www.science.org/doi/pdf/10.1126/science.aav1910}
}

@article{redekop2024fci,
	author = {Redekop, Evgeny and Zhang, Canxun and Park, Heonjoon and Cai, Jiaqi and Anderson, Eric and Sheekey, Owen and Arp, Trevor and Babikyan, Grigory and Salters, Samuel and Watanabe, Kenji and Taniguchi, Takashi and Huber, Martin E. and Xu, Xiaodong and Young, Andrea F.},
	date = {2024/11/01},
	doi = {10.1038/s41586-024-08153-x},
	id = {Redekop2024},
	isbn = {1476-4687},
	journal = {Nature},
	number = {8039},
	pages = {584--589},
	title = {{Direct magnetic imaging of fractional Chern insulators in twisted MoTe$_2$}},
	url = {https://doi.org/10.1038/s41586-024-08153-x},
	volume = {635},
	year = {2024}}

@article{zhou2022superconductivity-bernal,
author = {Haoxin Zhou  and Ludwig Holleis  and Yu Saito  and Liam Cohen  and William Huynh  and Caitlin L. Patterson  and Fangyuan Yang  and Takashi Taniguchi  and Kenji Watanabe  and Andrea F. Young },
title = {{Isospin magnetism and spin-polarized superconductivity in Bernal bilayer graphene}},
journal = {Science},
volume = {375},
number = {6582},
pages = {774-778},
year = {2022},
doi = {10.1126/science.abm8386},
URL = {https://www.science.org/doi/abs/10.1126/science.abm8386},
eprint = {https://www.science.org/doi/pdf/10.1126/science.abm8386}}

@article{zhou2021halfmetal,
	author = {Zhou, Haoxin and Xie, Tian and Ghazaryan, Areg and Holder, Tobias and Ehrets, James R. and Spanton, Eric M. and Taniguchi, Takashi and Watanabe, Kenji and Berg, Erez and Serbyn, Maksym and Young, Andrea F.},
	date = {2021/10/01},
	doi = {10.1038/s41586-021-03938-w},
	id = {Zhou2021},
	isbn = {1476-4687},
	journal = {Nature},
	number = {7881},
	pages = {429--433},
	title = {Half- and quarter-metals in rhombohedral trilayer graphene},
	url = {https://doi.org/10.1038/s41586-021-03938-w},
	volume = {598},
	year = {2021}}

@article{tang2020hubbard,
	author = {Tang, Yanhao and Li, Lizhong and Li, Tingxin and Xu, Yang and Liu, Song and Barmak, Katayun and Watanabe, Kenji and Taniguchi, Takashi and MacDonald, Allan H. and Shan, Jie and Mak, Kin Fai},
	date = {2020/03/01},
	doi = {10.1038/s41586-020-2085-3},
	id = {Tang2020},
	isbn = {1476-4687},
	journal = {Nature},
	number = {7799},
	pages = {353--358},
	title = {{Simulation of Hubbard model physics in WSe$_2$/WS$_2$ Moir{\'e} superlattices}},
	url = {https://doi.org/10.1038/s41586-020-2085-3},
	volume = {579},
	year = {2020}}

@article{xu2020correlated,
	author = {Xu, Yang and Liu, Song and Rhodes, Daniel A. and Watanabe, Kenji and Taniguchi, Takashi and Hone, James and Elser, Veit and Mak, Kin Fai and Shan, Jie},
	date = {2020/11/01},
	doi = {10.1038/s41586-020-2868-6},
	id = {Xu2020},
	isbn = {1476-4687},
	journal = {Nature},
	number = {7833},
	pages = {214--218},
	title = {{Correlated insulating states at fractional fillings of Moir{\'e} superlattices}},
	url = {https://doi.org/10.1038/s41586-020-2868-6},
	volume = {587},
	year = {2020}}

@article{lu2025qahe,
	author = {Lu, Zhengguang and Han, Tonghang and Yao, Yuxuan and Hadjri, Zach and Yang, Jixiang and Seo, Junseok and Shi, Lihan and Ye, Shenyong and Watanabe, Kenji and Taniguchi, Takashi and Ju, Long},
	date = {2025/01/01},
	doi = {10.1038/s41586-024-08470-1},
	id = {Lu2025},
	isbn = {1476-4687},
	journal = {Nature},
	number = {8048},
	pages = {1090--1095},
	title = {{Extended quantum anomalous Hall states in graphene/hBN Moir{\'e} superlattices}},
	url = {https://doi.org/10.1038/s41586-024-08470-1},
	volume = {637},
	year = {2025}}

@article{Guo2025Jan,
	author = {Guo, Yinjie and Pack, Jordan and Swann, Joshua and Holtzman, Luke and Cothrine, Matthew and Watanabe, Kenji and Taniguchi, Takashi and Mandrus, David G. and Barmak, Katayun and Hone, James and Millis, Andrew J. and Pasupathy, Abhay and Dean, Cory R.},
	date = {2025/01/01},
	doi = {10.1038/s41586-024-08381-1},
	id = {Guo2025},
	isbn = {1476-4687},
	journal = {Nature},
	number = {8047},
	pages = {839--845},
	title = {{Superconductivity in 5.0$\,^{\circ}$ twisted bilayer WSe$_2$}},
	url = {https://doi.org/10.1038/s41586-024-08381-1},
	volume = {637},
	year = {2025}}

@article{cai2023fqaheMoTe2,
	author = {Cai, Jiaqi and Anderson, Eric and Wang, Chong and Zhang, Xiaowei and Liu, Xiaoyu and Holtzmann, William and Zhang, Yinong and Fan, Fengren and Taniguchi, Takashi and Watanabe, Kenji and Ran, Ying and Cao, Ting and Fu, Liang and Xiao, Di and Yao, Wang and Xu, Xiaodong},
	date = {2023/10/01},
	doi = {10.1038/s41586-023-06289-w},
	id = {Cai2023},
	isbn = {1476-4687},
	journal = {Nature},
	number = {7981},
	pages = {63--68},
	title = {{Signatures of fractional quantum anomalous Hall states in twisted MoTe$_2$}},
	url = {https://doi.org/10.1038/s41586-023-06289-w},
	volume = {622},
	year = {2023}}

@article{lu2024fqaheGraphene,
	author = {Lu, Zhengguang and Han, Tonghang and Yao, Yuxuan and Reddy, Aidan P. and Yang, Jixiang and Seo, Junseok and Watanabe, Kenji and Taniguchi, Takashi and Fu, Liang and Ju, Long},
	date = {2024/02/01},
	doi = {10.1038/s41586-023-07010-7},
	id = {Lu2024},
	isbn = {1476-4687},
	journal = {Nature},
	number = {8000},
	pages = {759--764},
	title = {Fractional quantum anomalous Hall effect in multilayer graphene},
	url = {https://doi.org/10.1038/s41586-023-07010-7},
	volume = {626},
	year = {2024}}

@article{
KimScience2023,
author = {S. Y. Frank Zhao  and Xiaomeng Cui  and Pavel A. Volkov  and Hyobin Yoo  and Sangmin Lee  and Jules A. Gardener  and Austin J. Akey  and Rebecca Engelke  and Yuval Ronen  and Ruidan Zhong  and Genda Gu  and Stephan Plugge  and Tarun Tummuru  and Miyoung Kim  and Marcel Franz  and Jedediah H. Pixley  and Nicola Poccia  and Philip Kim },
title = {Time-reversal symmetry breaking superconductivity between twisted cuprate superconductors},
journal = {Science},
volume = {382},
number = {6677},
pages = {1422-1427},
year = {2023},
doi = {10.1126/science.abl8371},
URL = {https://www.science.org/doi/abs/10.1126/science.abl8371},
eprint = {https://www.science.org/doi/pdf/10.1126/science.abl8371}}

@article{
YacobyScience2010,
author = {R. T. Weitz  and M. T. Allen  and B. E. Feldman  and J. Martin  and A. Yacoby },
title = {Broken-Symmetry States in Doubly Gated Suspended Bilayer Graphene},
journal = {Science},
volume = {330},
number = {6005},
pages = {812-816},
year = {2010},
doi = {10.1126/science.1194988},
URL = {https://www.science.org/doi/abs/10.1126/science.1194988},
eprint = {https://www.science.org/doi/pdf/10.1126/science.1194988}}

@article{
NovoselovScience2011,
author = {A. S. Mayorov  and D. C. Elias  and M. Mucha-Kruczynski  and R. V. Gorbachev  and T. Tudorovskiy  and A. Zhukov  and S. V. Morozov  and M. I. Katsnelson  and V. I. Fal’ko and A. K. Geim  and K. S. Novoselov },
title = {Interaction-Driven Spectrum Reconstruction in Bilayer Graphene},
journal = {Science},
volume = {333},
number = {6044},
pages = {860-863},
year = {2011},
doi = {10.1126/science.1208683},
URL = {https://www.science.org/doi/abs/10.1126/science.1208683},
eprint = {https://www.science.org/doi/pdf/10.1126/science.1208683}}

\onecolumngrid
\newpage
\makeatletter 

\begin{center}
\textbf{\large Supplementary Material for: \\ ``\@title ''} \\[10pt]
Chun-Tse Li$^{1,2}$, Tzen Ong$^1$, Max Geier$^3$, Hsin Lin$^1$, and Liang Fu$^3$ \\
\textit{$^1$Institute of Physics, Academia Sinica, Taipei 115201, Taiwan}\\
\textit{$^2$Department of Electrical and Computer Engineering, University of Southern California, Los Angeles, California 90089, USA}\\
\textit{$^3$Department of Physics, Massachusetts Institute of Technology, Cambridge, MA-02139, USA}\\
\end{center}
\vspace{10pt}

\setcounter{figure}{0}
\setcounter{section}{0}
\setcounter{equation}{0}

\renewcommand{\thefigure}{S\@arabic\c@figure}
\makeatother

\appendix 

\section{Neural Quantum State Architecture}
\label{app:NN-Architecture}

\subsection{Neural Network Quantum States and Generalized Slater Determinants}

In this subsection, we describe in detail the architecture of our neural-network quantum state for the spin-polarized attractive Fermi gas.  A fundamental requirement for any fermionic wavefunction is \emph{antisymmetry} under particle exchange:

The simplest type of wavefunction that guarantees this property is the \textit{Slater determinant}, constructed by antisymmetrizing the single-particle orbitals $\{\phi_\mu(\mathbf x)\}_{\mu=1}^N$:
\begin{align}
\label{eq:slater}
\Psi_{\mathrm{SD}}(\mathbf X)
&=\frac{1}{\sqrt{N!}}
\det\bigl[\phi_\mu(\mathbf x_j)\bigr]_{j,\mu=1}^N  
\nonumber\\
&=\frac{1}{\sqrt{N!}}
\begin{vmatrix}
\phi_1(\mathbf x_1) & \phi_2(\mathbf x_1) & \cdots & \phi_N(\mathbf x_1)\\
\phi_1(\mathbf x_2) & \phi_2(\mathbf x_2) & \cdots & \phi_N(\mathbf x_2)\\
\vdots               & \vdots               & \ddots & \vdots\\
\phi_1(\mathbf x_N) & \phi_2(\mathbf x_N) & \cdots & \phi_N(\mathbf x_N)
\end{vmatrix}.
\end{align}
where $\mathbf X\coloneqq (\mathbf x_1,...,\mathbf x_N)$. Eq.~\eqref{eq:slater} describes a non-interacting Fermi gas and is variationally optimal only at the mean-field level.  Because of its simplicity, it often serves as the starting point for more advanced methods.

One direct route to incorporate correlations is to promote each orbital to a \textit{generalized orbital} that depends on the positions of all other particles~\cite{pfau2020ab}:
\begin{equation}
\Phi^k_\mu(\mathbf x_j;\{\mathbf x_{/j}\}),
\quad
\{\mathbf x_{/j}\} = \{\mathbf x_1,\dots,\mathbf x_N\}\setminus \mathbf x_j,
\end{equation}
and then form a linear combination of $N_{\det}$ such determinants:
\begin{equation}
\Psi(\mathbf X)
=\frac{1}{\sqrt{N!}}
\sum_{k=1}^{N_{\det}}\,
\det\bigl[\Phi^k_\mu(\mathbf x_j;\{\mathbf x_{/j}\})\bigr]_{j,\mu=1}^N.
\end{equation}
Here, $\mu$ indexes orbitals, $j$ indexes particles, and $k$ indexes determinants.  Although a single generalized determinant is, in principle, universal to represent a correlated many-body wavefunction~\cite{pfau2020ab}, it is often numerically challenging to learn; hence previous works use a small set of determinants to systematically improve accuracy.

In the traditional variational and diffusion Monte Carlo, the generalized orbital can be constructed by the Slater-Jastrow-Backflow wavefunction ansatz:
\begin{align}
    \Psi_{\text{SJB}}(\mathbf X)=J(\mathbf X)\det\bigl[\phi_\mu(\mathbf x_j+\bm\xi_j(\mathbf X))\bigr]_{j,\mu=1}^N
\end{align}
where $J(\mathbf X)$ is the Jastrow factor encoding dynamic electron–electron correlations, and $\bm\xi_j(\mathbf X)$ is the backflow displacement that “dresses” each electron coordinate with many-body screening effects. Despite its huge success, both $J(\cdot)$ and $\bm\xi_j(\cdot)$ must be chosen \textit{a priori}, often by a good physical intuition or trial and error. This hand-crafting wavefunction sometimes can miss important higher-order or long-range correlations and the limited number of parameters can further restrict the functional space of the many body wavefunction. In our work, each $\Phi^k_\mu$ is generated by a \textit{self-attention neural network} $g_\mu^k(\cdot)$:
\begin{equation}
\Phi^k_\mu(\mathbf x_j;\{\mathbf x_{/j}\})
= g^k_\mu\bigl(\mathbf X\bigr),
\end{equation}
where $g^k_\mu$ is required to be a \textit{permutation-equivariant} function such that
\begin{align}
    g_\mu^k(P_{j\mu}\mathbf X)=P_{j\mu} g^k_\mu(\mathbf X)=g_j^k(\mathbf X)
\end{align}
The $P_{j\mu}$ is the permutation operator that exchange the particle $j$ and $\mu$. The permutation equivarinace property thus ensure the antisymmetric property of a generalized Slater determinant: 
\begin{align}
    \det\bigl[g_\mu^k(P_{j'\mu'} \mathbf X) \bigr]_{j,\mu = 1}^N&=\det\bigl[P_{j'\mu'}g_{\mu}^k(\mathbf X)\bigr]_{j,\mu=1}^{N} =-\det\bigl[g_{\mu}^k(\mathbf X)\bigr]_{j,\mu=1}^N
\end{align}
This \textit{form-free} construction allows the network to discover both antisymmetric structure and many-body correlations \textit{ab initio}, without imposing Pfaffian, geminal, or backflow templates by hand.

\subsection{Self-Attention Neural Network}

Now, we discuss the detail construction of the self-attention neural network that generate the generalized orbitals $\Phi^k_\mu(\mathbf x_j;\{\mathbf x_{/j}\})$.

In the periodic system, one need to enforce the wavefunction to satisfy the periodic boundary condition (PBC):
\begin{align}
    \Psi(\mathbf x_1,..., \mathbf x_j+\mathbf L,...,\mathbf x_N) = \Psi(\mathbf x_1,...,\mathbf x_j,...,\mathbf x_N).
\end{align}
where $\mathbf L= n_1 \mathbf L_1+n _2\mathbf L_2$, and $n_1,n_2\in\Z$. The vectors $\mathbf L_1$ and $\mathbf L_2$ are the supercell vectors in the $x$ and $y$ direction. A direct way to enforce the PBCs is by passing the coordinate quantities into a periodic function; in our case, we consider the following feature map
\begin{align}
    \mathbf f^{(0)}=\Bigl[\bigl[\cos\bigl(\mathbf G_i\cdot\mathbf X\bigr)\bigl]_{i=1}^2, \bigl[\sin\bigl(\mathbf G_i\cdot \mathbf X\bigr)\bigr]_{i=1}^2\Bigr]
\end{align}
where we use the notation $[\ \cdot, \cdot \ ]$ to denote the concatenation, and $\mathbf G_i$ denotes the reciprocal lattice vector such that $\mathbf G_i \cdot \mathbf L_j=2\pi \delta_{ij}$. This ensures the the feature $\mathbf f_0$ is invariant under the translation of an integer multiple of supercell size $\mathbf L$. For the individual particle, the feature map is denoted as
\begin{align}
    \mathbf f^{(0)}_j=\Bigl[\bigl[\cos\bigl(\mathbf G_i\cdot\mathbf x_j\bigr)\bigl]_{i=1}^2, \bigl[\sin\bigl(\mathbf G_i\cdot \mathbf x_j\bigr)\bigr]_{i=1}^2\Bigr].
\end{align}
The initial feature is then mapped into a learnable hidden state $\mathbf h_0$ by a linear transformation
\begin{align}
    \mathbf h^{(0)}= W^{(0)} \mathbf f^{(0)}, \quad \mathbf h^{(0)}_j=W^{(0)}\mathbf f^{(0)}_j
\end{align}
where $W_0\in\R^{d_h\times 2d_{\dim}}$, and $d_{\dim}$ denotes the system dimension (in our case $d_{\dim}=2$) while $d_h$ denotes the hidden layer dimension.

The hidden states $\mathbf h^{(0)}_j$ are passed into $L$ layers of multi-head self-attention layer followed by the multi-layer perceptrons (MLP) to get the final output $\mathbf h^{(L)}$, where the high level graphical illustration can be found in Fig.~\ref{fig:Psiformer-Architecture}. The multi-head self-attention layer consists of three type of features -- query, key and value which can be obtained by a linear transformation on the hidden state:
\begin{align}
    \mathbf Q^{(\ell)}_{h,j}&=W_{Q,h}^{(\ell)}\mathbf h_j^{(\ell)}\in \R^{d_{\text{Attn}}}, 
    \nonumber \\[0.2cm] 
    \mathbf{K}_{h,j}^{(\ell)}&=W_{K,h}^{(\ell)}\mathbf h_j^{(\ell)}\in \R^{d_{\text{Attn}}}, 
    \nonumber \\[0.2cm] 
    \mathbf V_{h,j}^{(\ell)}&=W_{V,h}^{(\ell)}\mathbf h_j^{(\ell)}\in \R^{d_{\text{Attn,v}}}
\end{align}
where $W_Q^{(\ell)}, W_K^{(\ell)}\in \R^{d^{\text{Attn}}\times d_{h}}$, $W_V^{(\ell)}\in \R^{d_{\text{Attn},v}\times d_h}$ are the learnable linear transformation matrices and $j$ indexes the particles, $h$ indexes the heads and $\ell$ indexes the layers. We denote the concatenation of query, key and value in the particle dimension as
\begin{align}
    \mathbf Q_h^{(\ell)} &=\Bigl[\  \mathbf Q_{h,j}^{(\ell)} \ \Bigr]_{j=1}^{N} \in \R^{N\times d_{\text{Attn}}}, 
    \nonumber \\[0.2cm] 
    \mathbf K_h^{(\ell)}&=\Bigl[\  \mathbf K_{h,j}^{(\ell)} \ \Bigr]_{j=1}^{N}\in \R^{N\times d_{\text{Attn}}}, 
    \nonumber \\[0.2cm] 
    \mathbf V_h^{(\ell)}&=\Bigl[\  \mathbf V_{h,j}^{(\ell)} \ \Bigr]_{j=1}^{N} \in \R^{N\times d_{\text{Attn,v}}}.
\end{align}
The single head self-attention is therefore given by:
\begin{align}
    \text{Self-Attn}_h(\mathbf h^{(\ell)})=\text{Softmax}\Biggl(\frac{\mathbf Q^{(\ell)}_{h}(\mathbf K^{(\ell)}_{h})^T}{\sqrt{d_{\text{Attn}}}}\Biggr) \mathbf V_h^{(\ell)},
\end{align}
and the multi-head attention is obtained by stacking the $N_{\text{heads}}$ single-head self-attention and apply a learnable projection back to the dimension $d_h$:
\begin{align}
    \text{Self-Attn}(\mathbf h^{(\ell)})=W_o^{(\ell)}\Bigl[\text{Self-Attn}_h(\mathbf h^{(\ell)})\Bigr]_{h=1}^{N_{\text{heads}}},
\end{align}
where $W_o^{(\ell)}\in \R^{N_{d_h}\times N_{\text{heads}}d_{\text{Attn,v}}}$. The intermediate feature and hidden state in layer $\ell+1$ is given by a residual sum of the previous hidden state and the present feature in order to enhance the trainability of deep neural network:
\begin{align}
    \mathbf f^{(\ell+1)}&=\mathbf h^{(\ell)}+\text{Self-Attn}(\mathbf h^{(\ell)})\in \R^{N\times d_h} \\[0.2cm]
    \mathbf h^{(\ell+1)}&=\mathbf f^{(\ell)}+\tanh(W^{(l)}_{\rm P}\mathbf f^{(\ell+1)} + \mathbf b^{(l)}_{\rm P})\in \R^{N\times d_h}
\end{align}
where $W^{(l)}_{\rm P} \in \mathbb{R}^{d_h \times d_h}$ and $\mathbf b^{(l)}_{\rm P} \in \mathbb{R}^{d_h}$ are weights and biases of the MLP (here for a single layer) with $\tanh$ non-linearity.
After $L$ layers of multi-head attention and the MLP we project the last hidden state $\mathbf h^{(L)}$ to dimension $N$ to get $N_{\det}$ individual $N\times N$ orbital matrices:
\begin{align}
    [\Phi^k_\mu(\mathbf x_j;\{\mathbf x_{/j}\})]_{j,\mu=1}^{N}=W_k\mathbf h^{(L)}\in \C^{N\times N}
\end{align}
where $W_k\in \C^{N\times d_{h}}$, and $k=1,...,N_{\det}$. Therefore, the full many body wavefunction can be obtained by linear combination of $N_{\det}$ of generalized Slater determinants:
\begin{equation}
\Psi(\mathbf X)
=\frac{1}{\sqrt{N!}}
\sum_{k=1}^{N_{\det}}\,
\det\bigl[\Phi^k_\mu(\mathbf x_j;\{\mathbf x_{/j}\})\bigr]_{j,\mu=1}^N.
\end{equation}

\begin{figure}[ht!]
    \centering
    \begin{subfigure}[b]{0.495\textwidth}
        \includegraphics[width=\textwidth]{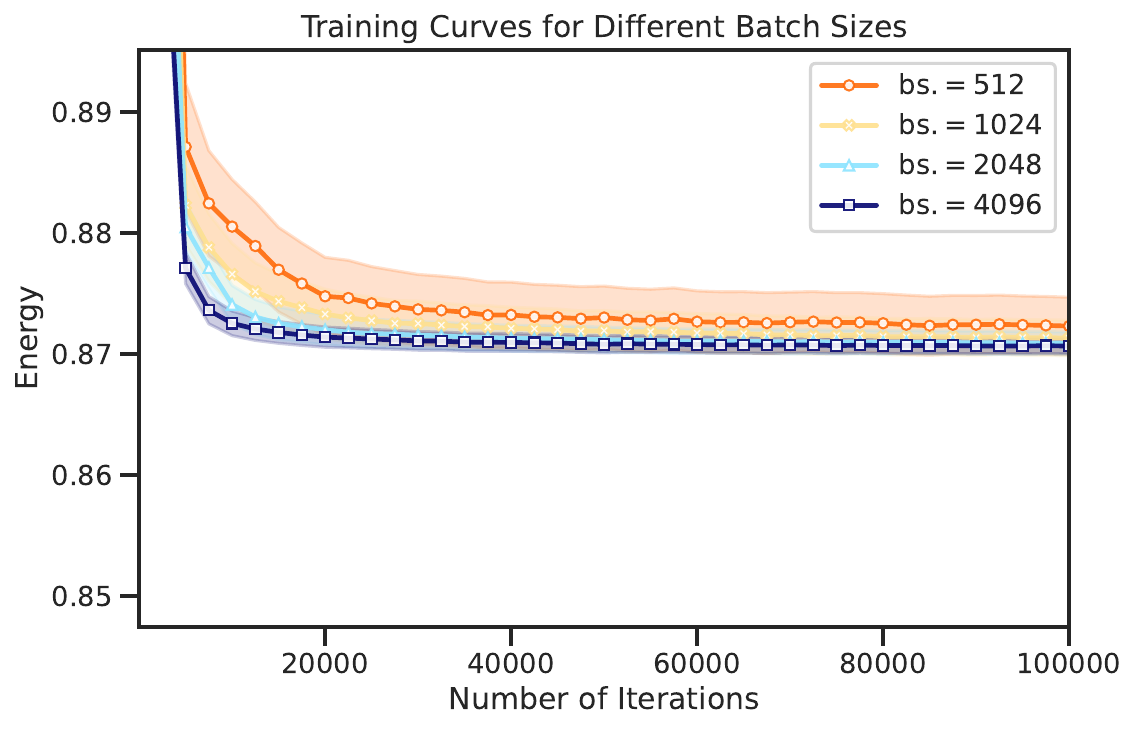}
    \end{subfigure}
    \hfill
    \begin{subfigure}[b]{0.495\textwidth}
        \includegraphics[width=\textwidth]{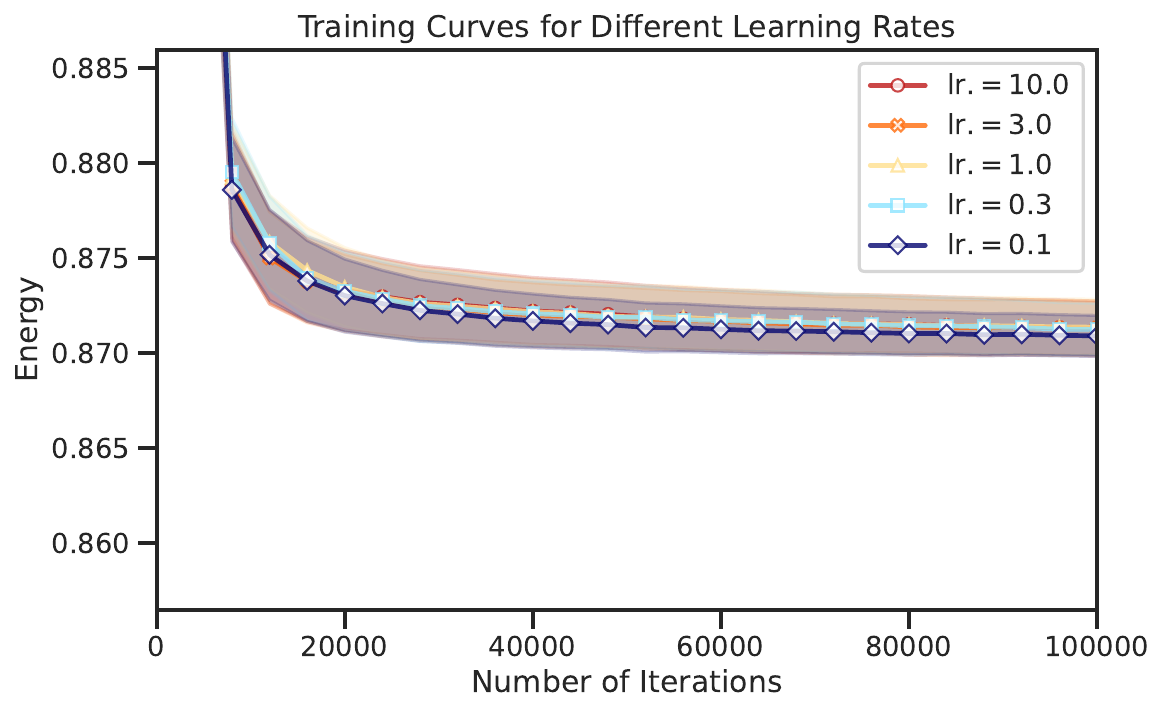}
    \end{subfigure}
    \begin{subfigure}[b]{0.495\textwidth}
        \includegraphics[width=\textwidth]{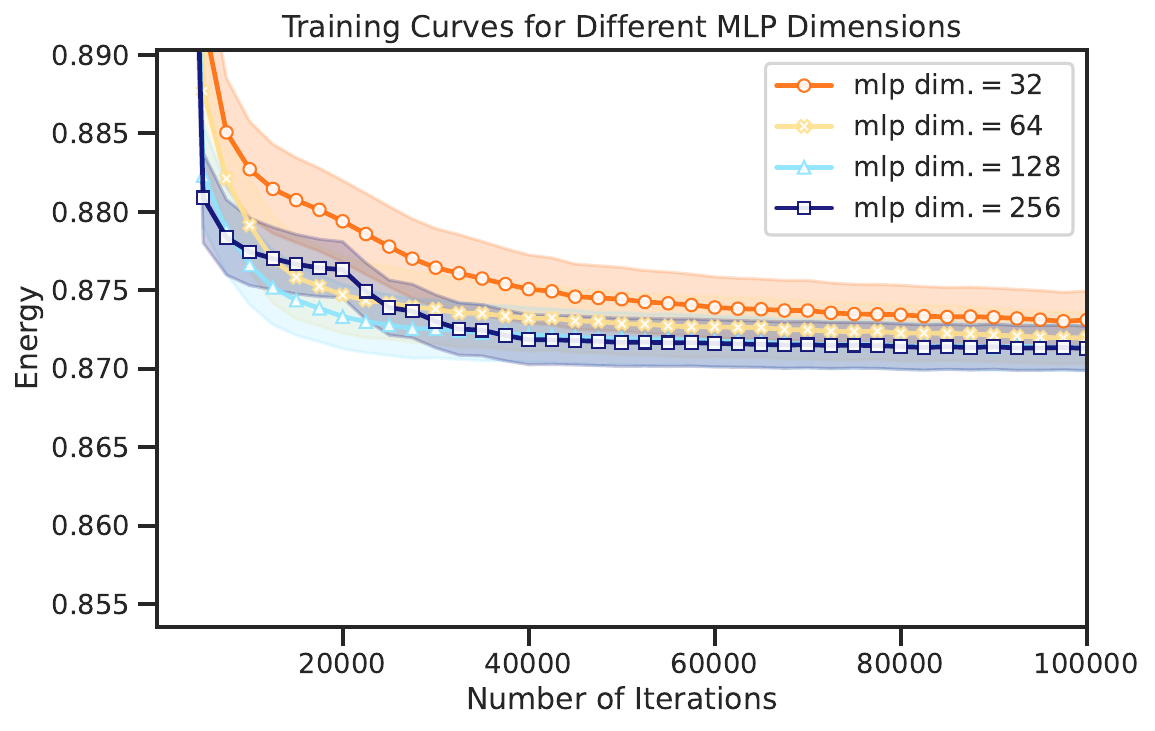}
    \end{subfigure}
    \hfill
    \begin{subfigure}[b]{0.495\textwidth}
        \includegraphics[width=\textwidth]{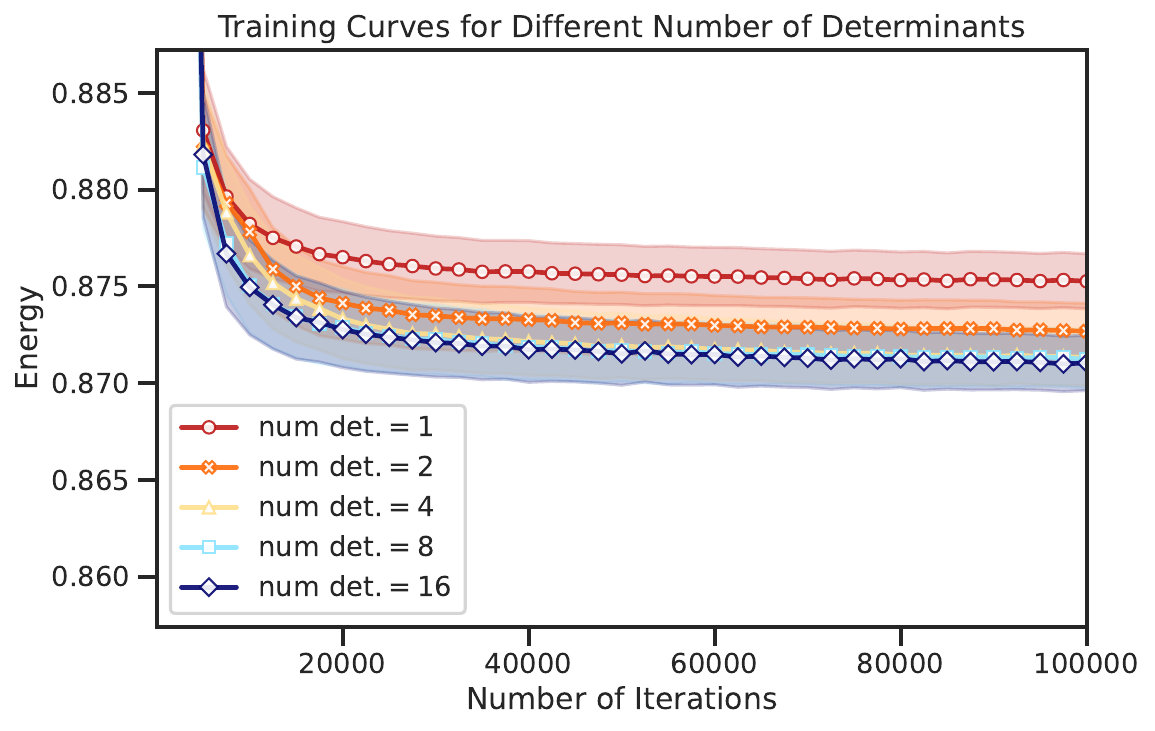}
    \end{subfigure}
    \caption{\justifying
    Training curves of the NNVMC ansatz under different optimization hyperparameters for the case $N=19$, $U=-10$, and $(L_x, L_y) = (30, 30)$. Each panel shows the mean energy (each point averaged over the past 2500 or 4000 iterations) as a function of the number of iterations (solid lines), together with the corresponding one–standard-deviation fluctuations (shaded regions). 
    \textbf{Top-left}: dependence on the batch size (512, 1024, 2048, 4096). 
    \textbf{Top-right}: dependence on the learning rate (10.0, 3.0, 1.0, 0.3, 0.1). 
    \textbf{Bottom-left}: dependence on the MLP hidden-layer dimension (32, 64, 128, 256). 
    \textbf{Bottom-right}: dependence on the number of determinants (1, 2, 4, 8, 16) in the Slater part of the wavefunction. 
    Across all hyperparameter choices, the optimization remains stable and converges to statistically consistent energies, demonstrating the robustness of both the architecture and the training procedure.
    }
    \label{fig:convergence-analysis}
\end{figure}

\textit{Hyperparameter Analysis.} Our NN results are robust and reproducible in a range of hyperparameters. We have performed a set of convergence studies for the hyperparameters, batch size $\in [512, 1024, 2048, 4096]$, learning rate $\in [10.0, 3.0, 1.0, 0.3, 0.1]$, MLP dimension $\in [32, 64, 128, 256]$ and No. of determinants $\in [1, 2, 4, 8, 16]$.
The detailed hyperparameters used in this work can be found in Table.~\ref{table:hyperparameters}, and the convergence study with respect to different hyperparameters is plotted in Fig.~\ref{fig:convergence-analysis}.

\begin{table}[t!]
  \caption{Table of hyperparameters of the self-attention neural network and VMC calculation.}
  \begin{ruledtabular}
  \begin{tabular}{llr}
     & \multicolumn{1}{c}{Parameter} & \multicolumn{1}{c}{Value} \\
    \hline
    \multirow{8}{*}{Architecture}
      & Network layers                 & 6 \\
      & Attention heads per layer      & 3 \\
      & Attention dimension (query, key) & 16 \\
      & Attention dimension (value)    & 16 \\
      & Perceptron dimension           & 128 \\
      & No.\ perceptrons per layer     & 1 \\
      & Layer norm                     & False \\
      & Determinants                   & 4 \\
    \cline{1-3} 

    \multirow{5}{*}{Training}
      & Training iterations            & 100000 \\
      & Learning rate schedule         & $\eta_0(1+t/t_0)^{-1}$ \\
      & Initial learning rate $\eta_0$ & 1 \\
      & Learning rate delay $t_0$      & $1\times10^{5}$ \\
      & Local energy clipping $\rho$   & 5.0 \\
    \cline{1-3}

    \multirow{3}{*}{MCMC}
      & Batch size                     & 1024 \\
      & Burn in steps                  & 200 \\ 
      & Sample move width              & 0.2 \\
    \cline{1-3}

    \multirow{2}{*}{KFAC}
      & Norm constraint                & $1\times10^{-3}$ \\
      & Damping                        & $1\times10^{-3}$ \\
  \end{tabular}
  \end{ruledtabular}
  \label{table:hyperparameters}
\end{table}

\begin{figure}[ht!]
    \centering
    \begin{subfigure}[b]{0.495\textwidth}
        \includegraphics[width=\textwidth]{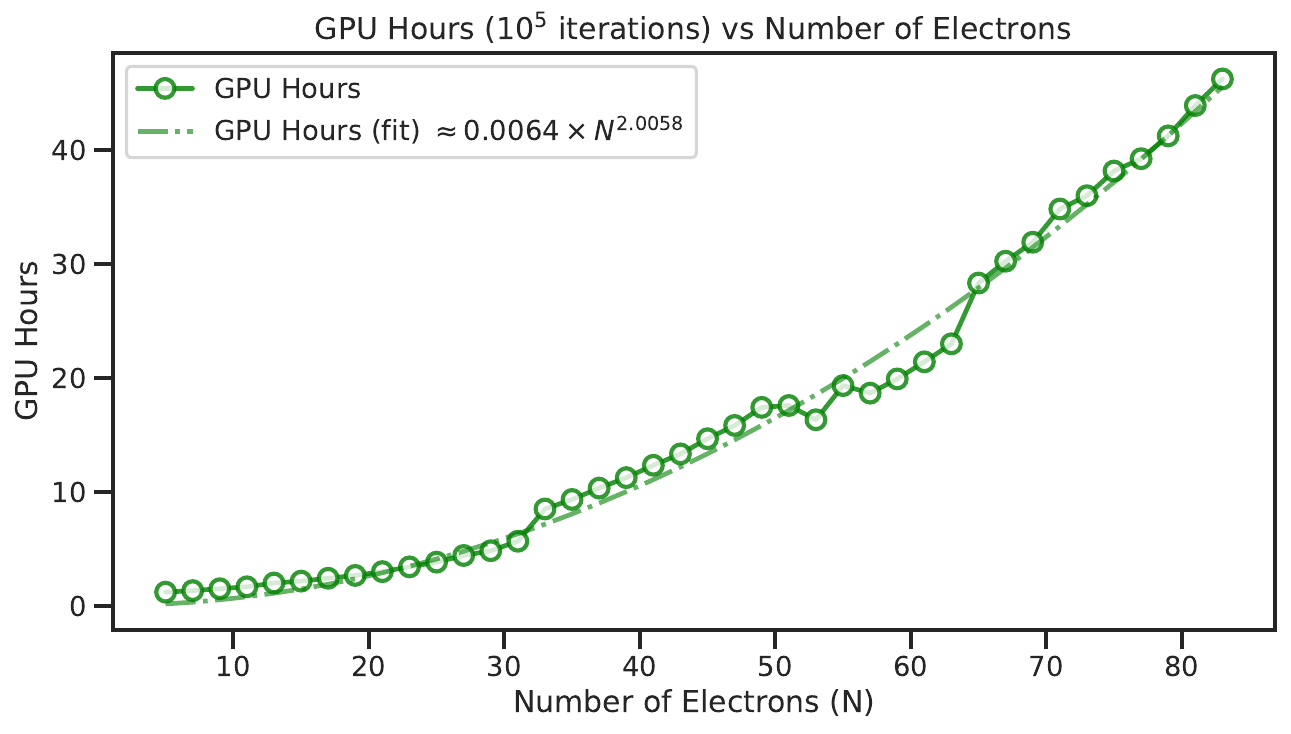}
    \end{subfigure}
    \hfill
    \begin{subfigure}[b]{0.495\textwidth}
        \includegraphics[width=\textwidth]{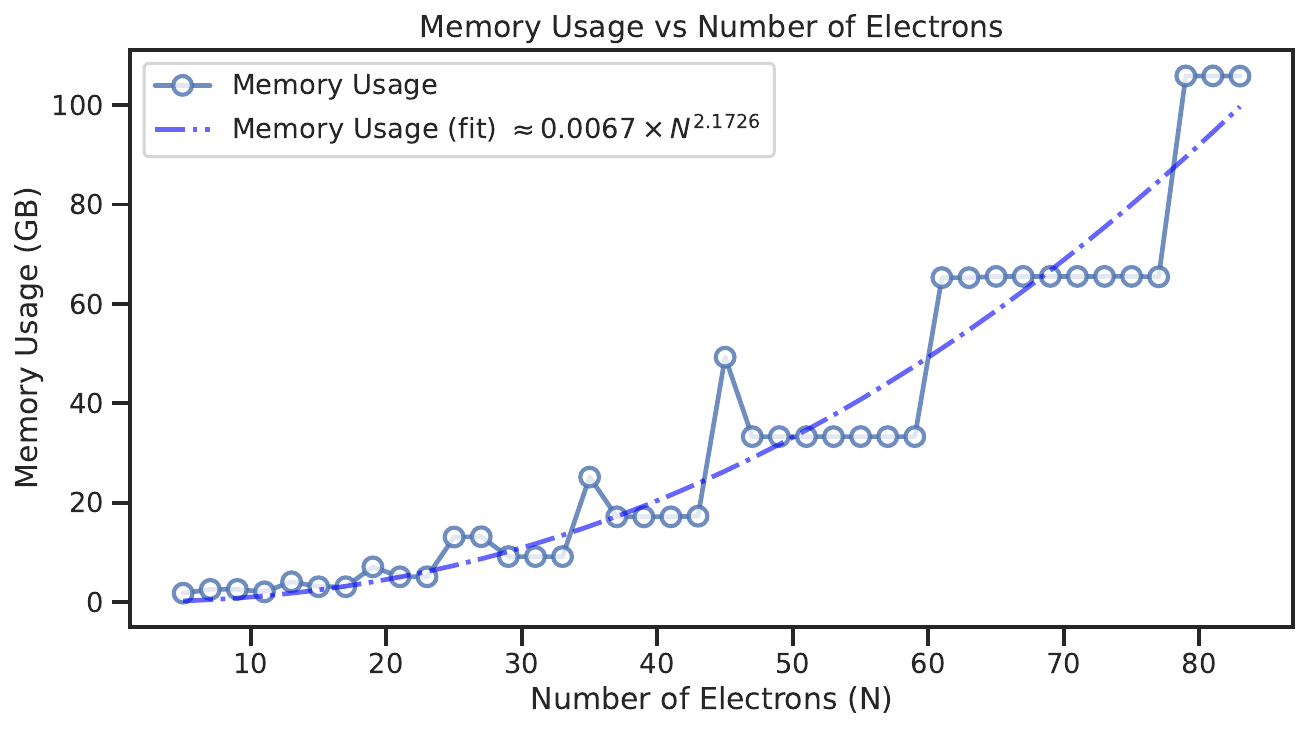}
    \end{subfigure}
    \caption{\justifying
    GPU runtime and memory scaling of the NNVMC method as a function of the number of electrons $N$. 
    \textbf{Left}: GPU hours required for $10^{5}$ training iterations. The data are well described by a power-law fit 
    $\mathrm{GPU\;hours} \approx 0.0064\times N^{2.0058}$, indicating an approximately quadratic growth with system size. 
    \textbf{Right}: Peak GPU memory consumption during training, which follows 
    $\mathrm{Memory} \approx 0.0067 \times N^{2.1726}$ (in GB). 
    Both scalings arise primarily from the quadratic complexity of the self-attention layers in the neural-network ansatz, 
    whose attention maps scale as $O(N^{2})$ in the number of particles. 
    Although architectural optimizations reduce prefactors, the overall quadratic behavior is intrinsic to attention-based 
    representations of many-electron wavefunctions.
    }
    \label{fig:gpu ram scaling}
\end{figure}

\textit{Computation overhead.}
For fixed network hyperparameters and Monte Carlo batch size, the dominant computational cost of our NNVMC ansatz scales quadratically with the number of particles $N$. In each transformer layer, building the self-attention maps and applying them to the value vectors costs $\mathcal{O}\big(L N_{\mathrm{heads}} (d_{\mathrm{attn}} + d_{\mathrm{val}}) N^{2}\big)$ operations, while projecting the final hidden representations to $N_{\mathrm{det}}$ sets of $N \times N$ orbitals contributes an additional $\mathcal{O}\!\big(N_{\mathrm{det}} d_{h} N^{2}\big)$ term.  Determinant updates and the evaluation of the two-body potential are likewise $\mathcal{O}(N^{2})$ per configuration, whereas embedding and MLP layers contribute only subleading $\mathcal{O}(N)$ work.  Overall, the cost of a single forward/backward pass and local-energy evaluation therefore scales as
\[
\mathcal{O}\Big(N^{2}\big[L N_{\mathrm{heads}} (d_{\mathrm{attn}} + d_{\mathrm{val}}) + N_{\mathrm{det}} d_{h}\big]\Big),
\]
consistent with the empirically observed quadratic dependence of runtime and memory on $N$ in Fig.~\ref{fig:gpu ram scaling}.

\subsection{Comparison to other numerical techniques}

In comparison to DMRG \cite{SchollwockRMP2005}, NN-VMC can operate on extended two-dimensional systems with equal extent in all directions -- a crucial requirement to accurately represent a Fermi gas. In comparison to full configuration interaction methods, NN-VMC operates in first-quantized continuum space and thereby avoids the need for truncating the full set of Slater determinants. Previous benchmarks \cite{geier2025attention} demonstrated that by doing so, the NN-VMC achieves a lower variational energy, i.e. a more optimal wavefunction.

\section{Projection onto angular momentum subspace}
\label{app:AM-projection}
In this section we describe how to project a candidate wavefunction onto the four one-dimensional irreducible representations of the $C_{4}$ point group. Given a variational wavefunction $\Psi(\mathbf X)$, define its images under successive $\pi/2$ rotations by the operator $\mathbf{R}_{\pi/2}$:
\begin{align}
    \Psi_k(\mathbf{X}) \equiv \Psi(\mathbf{R}^k_{\pi/2}\mathbf{X}),\qquad k=0,1,2,3.
\end{align}
Any such function admits a decomposition into components belonging to the four $C_{4}$ eigenspaces, labeled by $m=0,1,2,3$ with eigenvalue $e^{im\pi/2}$:
\begin{align}
    \Psi(\mathbf{X})=\sum_{m=0}^3 \tilde\Psi_m(\mathbf{X}),
\end{align}
where each component satisfies $\tilde\Psi_m(\mathbf{R}_{\pi/2}\mathbf{X})=e^{im\pi/2}\,\tilde\Psi_m(\mathbf{X})$. Stacking the rotated wavefunctions into a column vector,
\begin{align}
    \begin{pmatrix}
    \Psi_0 \\ \Psi_1 \\ \Psi_2 \\ \Psi_3
    \end{pmatrix}
    =
    M
    \begin{pmatrix}
    \tilde\Psi_0 \\ \tilde\Psi_1 \\ \tilde\Psi_2 \\ \tilde\Psi_3
    \end{pmatrix},
    \qquad
    M=
    \begin{pmatrix}
    1 & 1  & 1  & 1 \\
    1 & i  & -1 & -i \\
    1 & -1 & 1  & -1 \\
    1 & -i & -1 & i
    \end{pmatrix},
\end{align}
and inverting $M$ gives the components in each $C_4$ sector:
\begin{equation}
    \begin{pmatrix}
        \tilde\Psi_0 \\ \tilde\Psi_1 \\ \tilde\Psi_2 \\ \tilde\Psi_3
    \end{pmatrix}
    =
    \frac{1}{4}
    \begin{pmatrix}
        1 & 1  & 1 & 1 \\ 
        1 & -i & -1 & i \\
        1 & -1 & 1 & -1 \\
        1 & i  & -1 & -i 
    \end{pmatrix}
    \begin{pmatrix}
        \Psi_0 \\ \Psi_1 \\ \Psi_2 \\ \Psi_3
    \end{pmatrix}.
\end{equation}
Equivalently, in compact form,
\begin{align}
    \tilde\Psi_m=\frac{1}{4}\sum_{k=0}^3 e^{-ikm\pi/2}\,\Psi_k,
\end{align}
as stated in Eq.~\eqref{eq:AM-projection}.

\begin{figure}[ht!]
    \centering
    \includegraphics[width=0.75\linewidth]{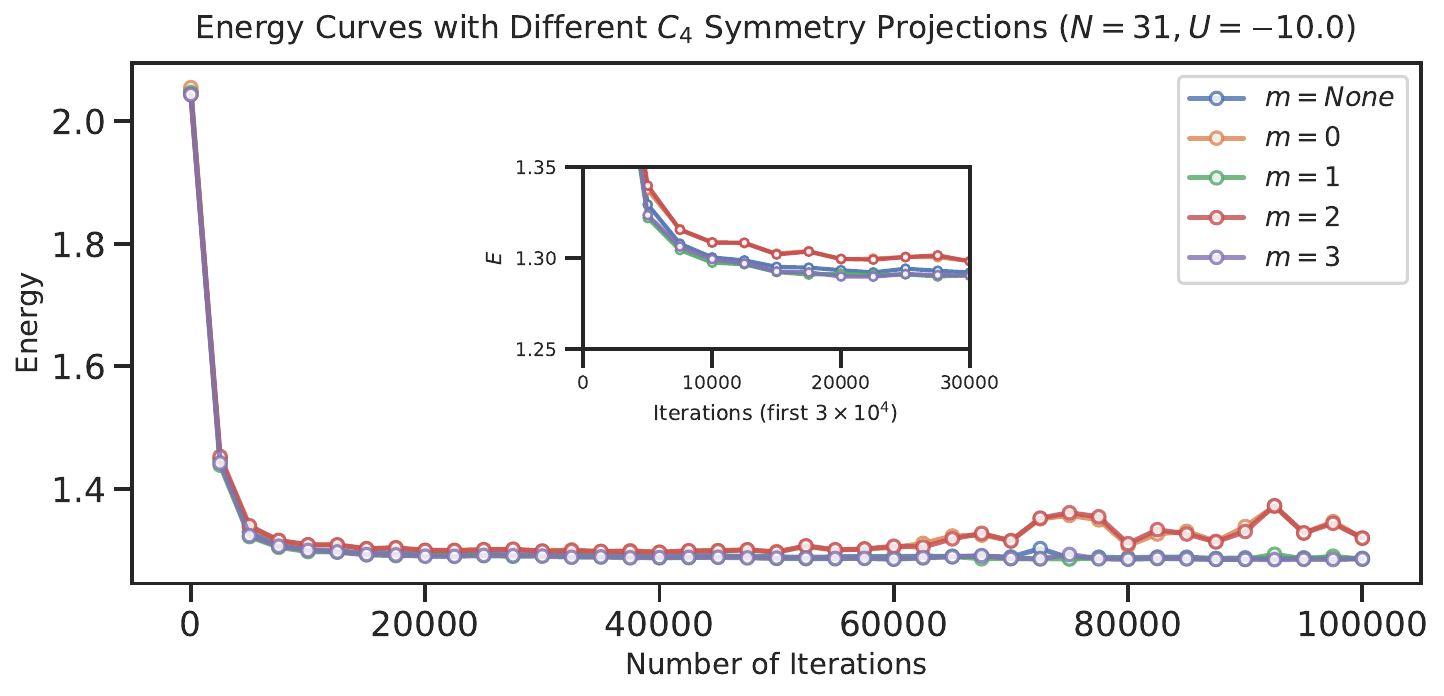}
    \caption{\justifying
    Energy convergence curves for different $C_{4}$ symmetry projections in the $N = 31$, $U = -10$ system. 
    The projection label $m$ denotes the total angular-momentum channel, with $m = \text{None}$ indicating that no symmetry projection is applied. 
    The inset shows a magnified view of the first $3 \times 10^{4}$ iterations, where the energies of the projected sectors begin to separate. 
    Among all channels, the $m = 1$ and $m = 3$ sectors yield the lowest variational energies, demonstrating that the time-reversal-symmetry–breaking states form the true ground-state manifold and that the NN wavefunction predominantly resides in the subspace spanned by the $m=1$ and $m=3$ channels.
    }
    \label{fig:C4 proj training}
\end{figure}

\begin{figure}[ht!]
    \centering
    \includegraphics[width=0.7\linewidth]{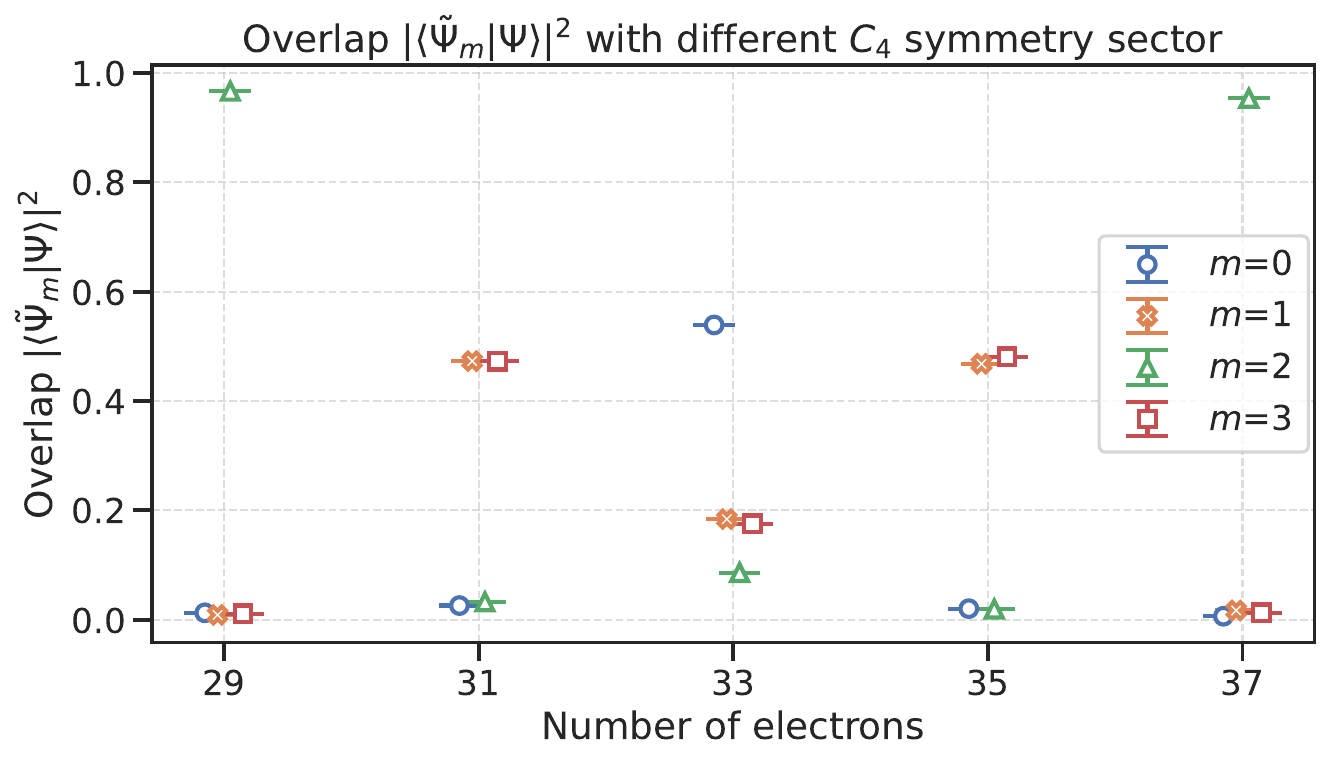}
    \caption{\justifying
    Overlap $|\langle \tilde{\Psi}_{m} \vert \Psi \rangle|^{2}$ between the variational NNVMC ground state $\Psi$ and the $C_{4}$ symmetry-projected states $\tilde{\Psi}_{m}$ for different number of particles ($N = 29, 31, 33, 35, 37$). 
    }
    \label{fig:C4-overlap}
\end{figure}

\textit{Time-reversal symmetry breaking and $C_{4}$ symmetry projection.}
As discussed in the main text, the $C_{4}$ symmetry projection is crucial for isolating the TRSB states with different chirality. In Fig.~\ref{fig:C4 proj training} we show the energy loss curves for $N = 31$, $U = -10.0$, and $(L_x,L_y) = (41.798, 41.798)$, where each state is projected into a different $C_{4}$ symmetry sector. For $N = 31$, the expected ground-state angular momenta are 
$m = \pm (N-1)/2 \bmod 4 = 1,3$. We see that already at very early training iterations, the $m = 0,2$ sectors have split from the $m = 1,3$ sectors and from the unprojected network, indicating that the NN wavefunction almost entirely lies in the manifold spanned by the $m = 1,3$ states. Moreover, in Fig.~\ref{fig:C4-overlap} we plot the overlap $|\langle \tilde{\Psi}_m | \Psi \rangle|^{2}$ for the four angular-momentum channels. For $N = 29, 33, 37$, the expected ground-state angular momenta are $m = 2, 0, 2$, respectively, and indeed a large fraction of the unprojected wave function weight lies in these symmetry sectors. In contrast, for $N = 31, 35$, the expected ground states have $m = \pm 1$, corresponding to chiral $p_x + i p_y$ and $p_x - i p_y$ respectively; in these two cases, the NN learns an almost equal superposition of these two symmetry sectors, consistent with the discussion in the main text.

\section{Variational Monte Carlo}
\label{app:VMC}

\subsection{Wavefunction Optimization}
In variational Monte Carlo (VMC) the variational energy of a parametrized wave‑function
$\Psi_{\bm\theta}(\mathbf X)\in\mathbb C$ is
\begin{align}
    E(\bm\theta)
    =\frac{\displaystyle\int d\mathbf X \Psi_{\bm\theta}^*(\mathbf X)H\Psi_{\bm\theta}(\mathbf X)}{\displaystyle\int d\mathbf X|\Psi_{\bm\theta}(\mathbf X)|^{2}},
    \label{eq:vmc-energy}
\end{align}
where $\mathbf X=(\mathbf x_{1},\dots,\mathbf x_{N})$ collects all particle coordinates and
$H$ is the Hamiltonian.  
Allowing $\Psi_{\bm\theta}$ to be complex accommodates possible
time‑reversal–symmetry breaking.

Rewriting Eq.~\eqref{eq:vmc-energy} in terms of the probability density
\begin{align}
    p_{\bm\theta}(\mathbf X)
    &=\frac{|\Psi_{\bm\theta}(\mathbf X)|^{2}}
            {\displaystyle\int d\mathbf X |\Psi_{\bm\theta}(\mathbf X)|^{2}},
    \quad \quad E_L(\mathbf X)
    =\frac{H\,\Psi_{\bm\theta}(\mathbf X)}{\Psi_{\bm\theta}(\mathbf X)},
    \label{eq:local-energy}
\end{align}
where $E_L(\mathbf X)$ is the local energy. This yields the expectation form
\begin{align}
    E(\bm\theta)=
    \mathbb E_{\mathbf X\sim p_{\bm\theta}}\!\bigl[E_{\mathrm{L}}(\mathbf X)\bigr].
\end{align}
We can sample the particle configurations $\{\mathbf X^{(i)}\}_{i=1}^{N_s}$ from
$p_{\bm\theta}(\mathbf X)$ using the Metropolis–Hastings algorithm (even when the underlying wavefunction $\Psi_{\bm\theta}(\mathbf{X})$ is unnormalized), and the expectation value can be empirically estimated by the Monte Carlo average
\begin{align}
    E(\bm\theta)\;\approx\;
    \frac{1}{N_s}\sum_{i=1}^{N_s}E_{L}\bigl(\mathbf X^{(i)}\bigr),
\end{align}
where $N_s$ is the number of Markov‑chain samples. The gradient of the energy expectation can be directly calculated from the local energy and is given by the following formula~\cite{becca2017quantum}:
\begin{align*}
    g_a&=\partial_{\theta_a}E(\bm\theta)=2\mathbb{E}_{\mathbf X\sim p_{\bm\theta}}\bigl[(E_L(\mathbf X)-E(\bm\theta))O_a\bigr] \\[0.2cm]
    O_a&= \partial_{\theta_a}\log\Psi_{\bm\theta}(\mathbf X)
\end{align*}
and the local energy $E_L(\mathbf X)$ can be calculated from the log-wavefunction identity
\begin{align}
    E_{\text{kin}}(\mathbf X)
    &= -\frac{1}{2m}\sum_{j=1}^N\sum_{i=1}^{d_{\dim}}
       \left[
          \left(\frac{\partial \log \Psi_{\bm\theta}}{\partial x_{j,i}}\right)^{\!2}
          + \frac{\partial^{2} \log \Psi_{\bm\theta}}{\partial x_{j,i}^{2}}
       \right], \\[0.2cm]
    E_L(\mathbf X) &= E_{\text{kin}}(\mathbf X) + V(\mathbf X),
\end{align}
where the first index $j$ labels particles and the second index $i$ labels Cartesian components.

\subsection{Natural‑Gradient (Stochastic Reconfiguration)}

Direct stochastic‑gradient descent converges slowly because the energy landscape is highly anisotropic in parameter space.  
\textit{Stochastic reconfiguration} (SR) preconditions the gradient $g_a$ with the quantum geometric tensor (QGT):
\begin{align}
    S_{ab}=\mathbb E_{\mathbf X\sim p_{\bm \theta}} \bigl[O^*_{a}O_{b}\bigr]-\mathbb E_{\mathbf X\sim p_{\bm\theta}}\bigl[O^*_{a}\bigr]\mathbb E_{\mathbf X\sim p_{\bm\theta}}\bigl[O_{b}\bigr],
\end{align}
producing the natural gradient  
\begin{align}
\Delta\boldsymbol\theta = -\alpha\,S^{-1}\mathbf g,\label{eq:natgrad}
\end{align}
where $\alpha$ is the learning rate. The natural gradient follows the steepest descent direction in wavefunction space rather than parameter space \cite{sorella1998green,amari1998natural} and has already shown to be equivalent to performing the imaginary time evolution on the parameter manifold~\cite{stokes2020quantum} in the limit of $\alpha\to0$. An approximation to the QGT by considering the absolute value of $\Psi_{\bm\theta}(\mathbf X)$ is the Fisher information metric (FIM)
\begin{align}
    F_{ab}&=\mathbb E_{\mathbf X\sim p_{\bm\theta}}\bigl[\partial_{\theta_a}\log p_{\bm\theta}(\mathbf X) \partial_{\theta_b}\log p_{\bm{\theta}}(\mathbf X)\bigr] 
    =4\mathbb E_{\mathbf X\sim p_{\bm\theta}}\bigl[\tilde{O}_a\tilde{O}_b\bigr]
\end{align}
where $\tilde{O}_a=\partial_{\theta_a}\log|\Psi_{\bm\theta}(\mathbf X)|$. The FIM can be derived by calculating the QGT of the absolute wavefunction $|\Psi_{\bm\theta}(\mathbf X)|$ (up to constant scaling 4) and the Berry connection term $\mathbb E_{\mathbf X\sim p_{\bm\theta}}\bigl[O_a\bigr]$ vanishes when considering the absolute wavefunction.
Unfortunately, in modern NQS, both the dimension of $S$ and $F$ can exceed $10^{5}\!\times\!10^{5}$, rendering exact inversion impractical.

\subsection{Kronecker‑Factored Approximate Curvature (KFAC)}

To overcome this bottleneck we use the \emph{Kronecker‑factored Approximate Curvature} (KFAC) optimizer \cite{martens2015optimizing}, an efficient approximation to the natural gradient widely adopted in deep‑learning and, more recently, in VMC \cite{pfau2020ab}. The KFAC formulation assumes that the matrix element $F_{ab}\approx0$ if $\theta_a$ and $\theta_b$ are from different neural network layers. Therefore, the FIM is reduced to the direct sum of FIM blocks $F=\bigoplus_{\ell=1}^LF_{\ell}$ where $\ell$ labels the neural network layers.
Moreover, for each neural network layer, KFAC assumes that the FIM block factorizes as a Kronecker product of two much smaller matrices,
\begin{align}
F_{\ell}\approx A_\ell\otimes B_\ell,\label{eq:kfac}
\end{align}
where $A_\ell=\mathbb E_{\mathbf X\sim p_{\bm\theta}} \bigl[\bm\theta_{\ell} \bm\theta_{\ell}^T\bigr]$ is the covariance of the layer’s inputs and $B_\ell=\mathbb E_{\mathbf X\sim p_{\bm\theta}}\bigl[\bm g_\ell \bm g_\ell^T\bigr]$ is the covariance of the corresponding output gradients.  
Because $(A\otimes B)^{-1}=A^{-1}\otimes B^{-1}$, the inverse in Eq.~\eqref{eq:kfac} reduces to inverting two $\mathcal O(d)\times \mathcal O(d)$ matrices instead of a single $\mathcal {O}(d^{2}) \times \mathcal O(d^2)$ block, cutting both memory and computational cost from $\mathcal {O}(d^{4})$ to $\mathcal {O}(d^2)$ per layer.

\section{Supplementary Plots}
\label{app:supplementary-plots}


\begin{figure}[ht!]
    \centering
    \includegraphics[width=0.7\linewidth]{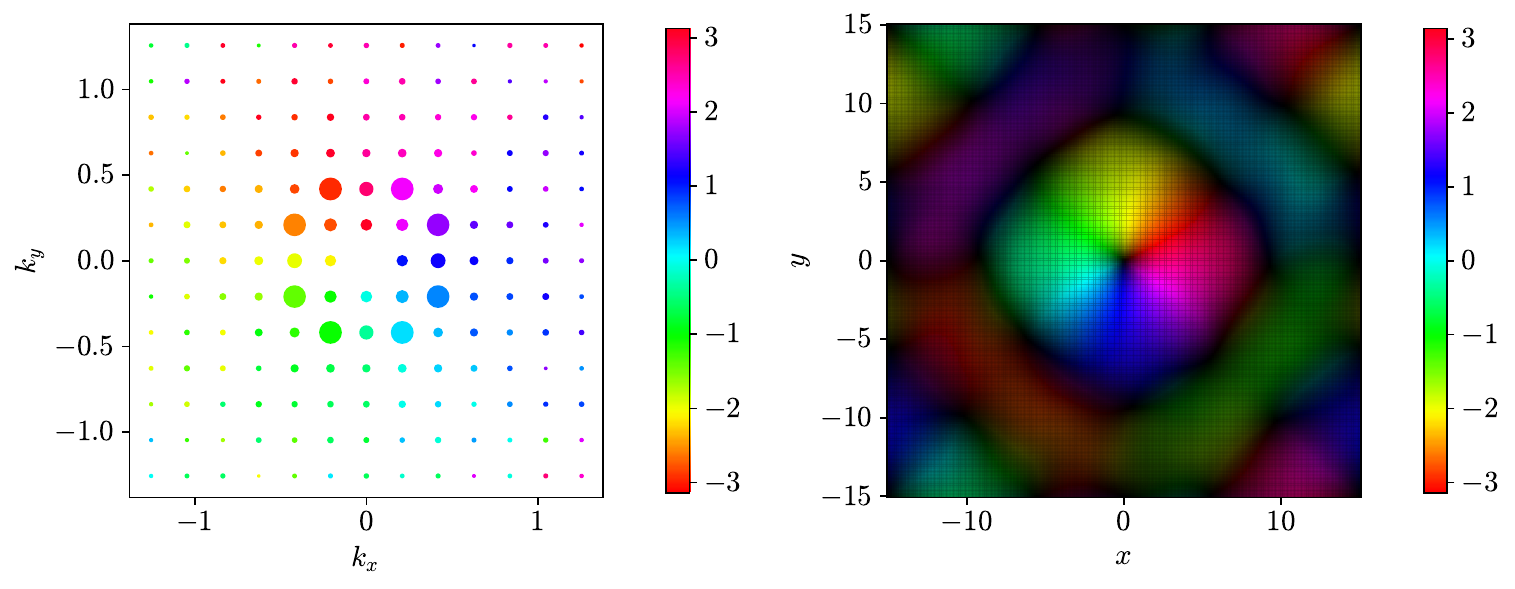}
    \caption{\justifying Leading eigenvector of the two-body reduced density matrix (2-RDM) in the $Q=0$ sector for $N=19$. (Left) Momentum-space representation $\Phi_0(\mathbf k)$, where marker size indicates $|\Phi_0(\mathbf k)|$ and color encodes the phase $\arg\Phi_0(\mathbf k)$. (Right) Real-space representation obtained by Fourier transforming $\Phi_0(\mathbf k)$; brightness indicates $|\Phi_0(\mathbf r)|$ and color encodes its phase. The $2\pi$ winding of the phase around the origin is consistent with chiral $p_x+i p_y$ pairing symmetry.}

    \label{fig:2-rdm-N=19}
\end{figure}

\begin{figure}[ht!]
    \centering
    \includegraphics[width=0.7\linewidth]{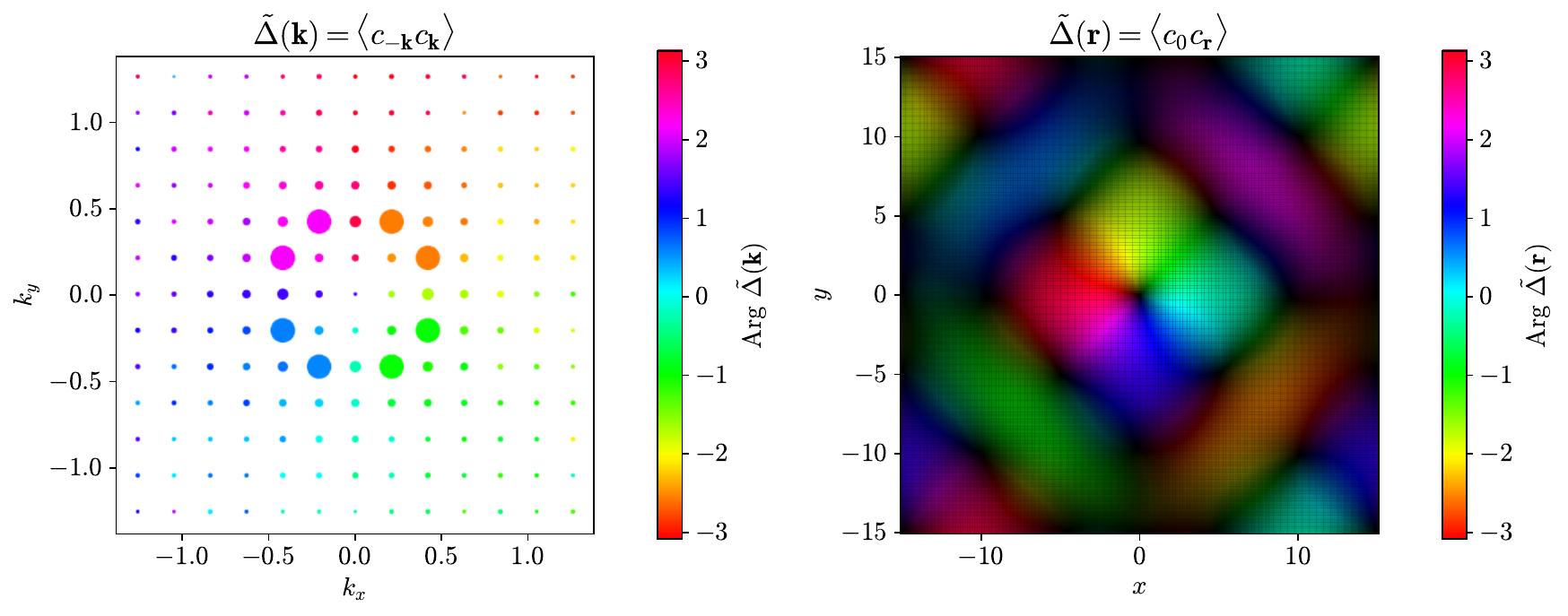}
    \caption{\justifying 
    Superconducting order parameter for $N=15$ at $U=-10$, $L=30$. The $N$- and $N+2=17$-particle wavefunctions are projected onto the $m=3$ and $m=0$ angular-momentum sectors, respectively. 
    (left) Cooper-pair amplitude $\tilde{\Delta}(\mathbf{k})=\langle \Psi_{N}\vert \hat c_{-\mathbf k}\hat c_{\mathbf k}\vert\Psi_{N+2}\rangle$ at the discrete momenta of the periodic box. Marker area scales with $|\tilde{\Delta}(\mathbf{k})|$, and color encodes the phase $\mathrm{Arg}\,\tilde{\Delta}(\mathbf{k})$, with a full color cycle corresponding to a $2\pi$ winding. 
    (right) Real-space representation obtained by Fourier transforming $\tilde\Delta(\mathbf k)$. The color indicates the phase $\mathrm{Arg}\,\tilde{\Delta}(\mathbf r)$, while brightness encodes the magnitude $|\tilde{\Delta}(\mathbf r)|$.}
    \label{fig:SC-order-parameter}
\end{figure}

{\it Superconducting order parameter.---} 
As an alternative to diagnosing pairing symmetry via the 2-RDM eigensystem, we adopt the formalism of Ref.~\cite{lu2010superconducting}, originally derived in Ref.~\cite{schrieffer2018theory}. 
Because the anomalous expectation value $\langle \hat c^\dagger_{\mathbf k}\hat c^\dagger_{-\mathbf k}\rangle$ vanishes in a fixed-particle-number wavefunction, we instead define the Cooper-pair amplitude through an off-diagonal matrix element between ground states differing by two fermions,
\begin{align}
\tilde{\Delta}(\mathbf k)=\langle \Psi_{N}\vert \hat c_{-\mathbf k}\hat c_{\mathbf k}\vert\Psi_{N+2}\rangle,
\label{eq:Delta_tilde}
\end{align}
where $|\Psi_{N}\rangle$ is the variational ground state in the $N$-particle sector.  

The resulting pair amplitude $\tilde \Delta (\mathbf k)$ is shown in Fig.~\ref{fig:SC-order-parameter}. 
Its phase winds by $2\pi$ as $\mathbf k$ encircles the Fermi surface, signaling a chiral $p_x+i p_y$ order parameter. 
Moreover, $|\tilde{\Delta}(\mathbf k)|$ remains finite at every Fermi momentum, consistent with a fully gapped spectrum. 
As required by Fermi statistics, the amplitude is odd under inversion, $\tilde{\Delta}(-\mathbf k)=-\tilde{\Delta}(\mathbf k)$; the small nonzero value at $\mathbf k=0$ arises from Monte Carlo noise.  
Importantly, neither pairing symmetry nor broken time-reversal symmetry is imposed in the ansatz—both emerge spontaneously from the variational optimization, underscoring the ability of the self-attention NQS to learn topological superconductivity \emph{ab initio}.

\begin{figure*}
    \centering
    \includegraphics[width=\linewidth]{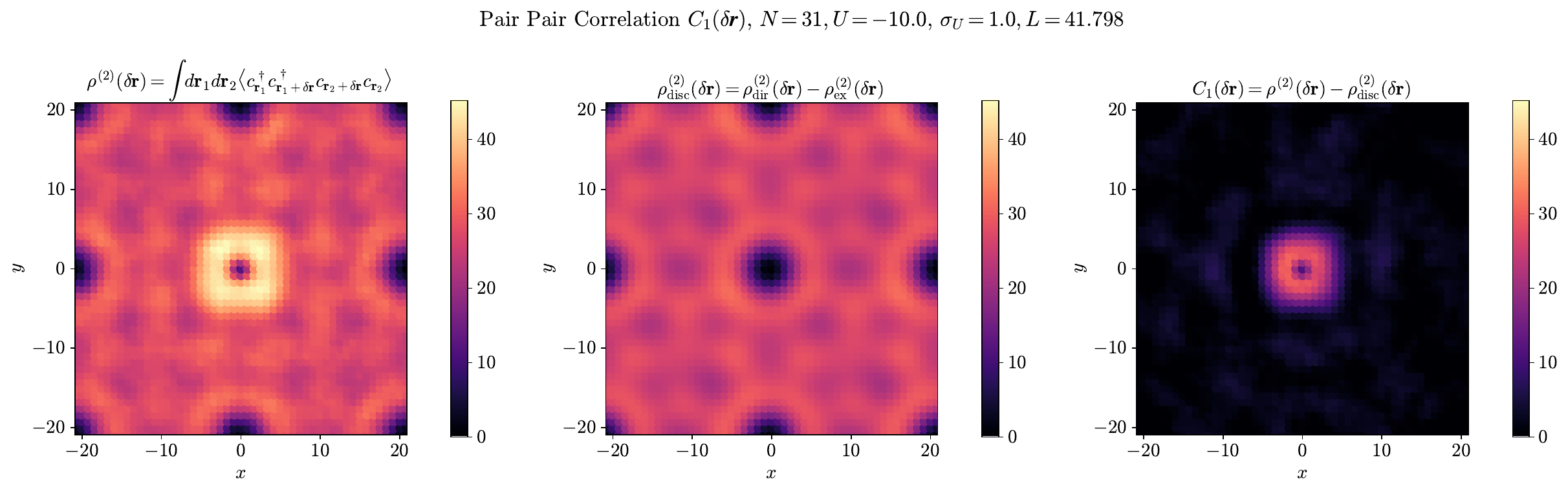}
    \caption{\justifying 
    Real-space pair correlator $C_1(\delta \mathbf r)$ for $N=31$, $U=-10$, $\sigma_U=1.0$, and $L=41.798$, with the wavefunction projected to the $m=3$ angular-momentum sector. 
    Left: the full two-body quantity $\rho^{(2)}(\delta\mathbf r)$. 
    Middle: the disconnected contribution $\rho^{(2)}_{\mathrm{disc}}(\delta\mathbf r)$. 
    Right: the connected correlator $C_1(\delta\mathbf r)=\rho^{(2)}(\delta\mathbf r)-\rho^{(2)}_{\mathrm{disc}}(\delta\mathbf r)$. 
    The bright annulus at $|\delta\mathbf r|\!\approx\!\xi_0$ identifies the characteristic Cooper-pair size $\xi_0$. 
    The node at $\delta\mathbf r=0$ in all panels reflects Pauli exclusion, while the vanishing of $\rho^{(2)}(\delta\mathbf r)$ on the boundary arises because $\hat\Delta_1(\delta\mathbf r)$ is odd under inversion, so points with $\delta\mathbf r\equiv -\delta\mathbf r$ on the torus yield $C_1(\delta\mathbf r)=0$.}
    \label{fig:C1-correlator}
\end{figure*}

\begin{figure*}
    \centering
    \includegraphics[width=0.6\linewidth]{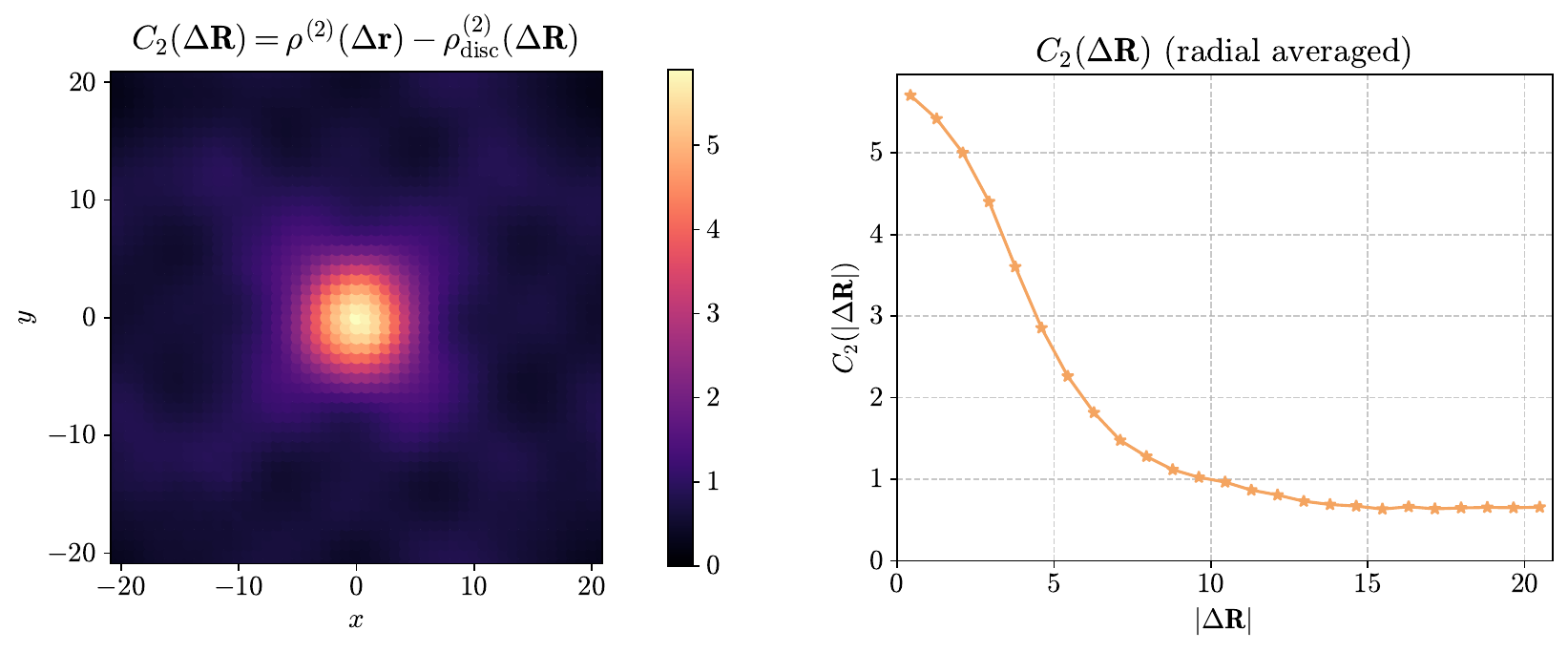}
    \caption{\justifying Connected pair correlator $C_2(\Delta\mathbf R)$ evaluated with the integration restricted to the disc $\mathcal D=\{|\mathbf r_1-\mathbf r_2|<\xi_0\}$. (Left) Real-space map of $C_2(\Delta\mathbf R)$. (Right) Radial average showing $C_2(|\Delta\mathbf R|)$ as a function of separation. The nonzero plateau for $|\Delta\mathbf R|\gg\xi_0$ demonstrates off-diagonal long-range order, whereas for a non-interacting Fermi gas one has $C_2(\Delta\mathbf R)\equiv 0$ for all $\Delta\mathbf R$.}

    \label{fig:C2-correlator}
\end{figure*}

\emph{Pair–pair correlators.} 
As a complement to the 2-RDM eigensystem discussed in the main text, we also consider 2-RDM–derived correlators that provide a more direct visualization of pairing. We decompose $\rho^{(2)}$ into a \emph{disconnected} contribution built from one-body correlators and a \emph{connected} remainder:
\begin{align}
    \rho^{(2)}_{\mathrm{disc}}(\mathbf r_1,\mathbf r_2;\mathbf r_1',\mathbf r_2')
    &= \langle \hat c^\dagger_{\mathbf r_1}\hat c_{\mathbf r_1'}\rangle \langle \hat c^\dagger_{\mathbf r_2}\hat c_{\mathbf r_2'}\rangle
     - \langle \hat c^\dagger_{\mathbf r_1}\hat c_{\mathbf r_2'}\rangle \langle \hat c^\dagger_{\mathbf r_2}\hat c_{\mathbf r_1'}\rangle, \\[0.2cm]
    \rho^{(2)}_{\mathrm{conn}} 
    &= \rho^{(2)}-\rho^{(2)}_{\mathrm{disc}}.
\end{align}
For a non-interacting (normal) Fermi gas, Wick factorization is exact and hence $\rho^{(2)}_{\mathrm{conn}}\equiv 0$.

Although $\rho^{(2)}$ fully encodes pair correlations, it is a high-dimensional object. To distill superconducting signatures into a more accessible form, we consider two derived correlators: the pair-size correlator $C_1(\delta\mathbf r)$, which characterizes the Cooper-pair size, and the off-diagonal long-range order correlator $C_2(\Delta\mathbf R)$, which probes long-range phase coherence.

\emph{Pair correlation length.---} The size of the Cooper pairs
can be extracted from the connected pair correlator,
We use the connected pair-separation correlator
\begin{align}
C_1(\delta \bm r)
&= \rho^{(2)}(\delta\mathbf r)\;-\;\rho_{\text{disc}}^{(2)}(\delta\mathbf r),
\end{align}
where
\begin{align}
\rho^{(2)}(\delta\mathbf r)
&= \big\langle \hat\Delta_1^\dagger(\delta\mathbf r)\,\hat\Delta_1(\delta\mathbf r)\big\rangle,\qquad
\hat\Delta_1(\delta \mathbf r)=\int_{\mathcal A} d\mathbf r \; \hat c_{\mathbf r+\delta \mathbf r}\,\hat c_{\mathbf r}.
\end{align}
Subtracting the disconnected terms in the second and third line ensures that $C_1(\delta\mathbf r) = 0$ for a non-interacting system because the disconnected terms are the Wick decomposition of the first term. By integrating over the entire supercell of area $\mathcal A= \det(L_1,L_2)$, $C_1(\delta\mathbf r)$ is signicant only for distances $\delta\mathbf r$ smaller than the pair correlation length $\xi_0$. At larger distances, uncorrelated contributions average to zero. Our numerical results, Fig.~\ref{fig:C1-correlator}, indicate a pair correlation length of the order of the interparticle distance.

\emph{Off-diagonal long-range order.---} 
A defining feature of a superconducting state is that the amplitude for creating a pair at one location and annihilating it at another remains finite even as the separation $\Delta\mathbf R \to \infty$ between their centers of mass becomes large. This is quantified by the connected pair correlator
\begin{align}
C_2(\Delta \bm R) &= \rho^{(2)}(\Delta\mathbf R)\;-\;\rho^{(2)}_{\text{disc}}(\Delta\mathbf R),
\end{align}
where the unsubtracted quantity is
\begin{align}
\rho^{(2)}(\Delta \mathbf R)
&= \int_{\cal D} d\bm r_1\, d\bm r_2\;
   \big\langle \hat c^\dagger_{\bm r_1+\Delta \bm R}\,
               \hat c^\dagger_{\bm r_2+\Delta \bm R}\,
               \hat c_{\bm r_2}\,
               \hat c_{\bm r_1} \big\rangle ,
\end{align}
and $\rho^{(2)}_{\text{disc}}$ denotes the disconnected (Wick) contribution, built from products of one-body correlators at the same density and boundary conditions. The integration domain ${\cal D}=\{(\bm r_1,\bm r_2):|\bm r_1-\bm r_2|<\xi_0\}$ restricts the internal pair separation so as to emphasize pairs of size $\lesssim\xi_0$. Fig.~\ref{fig:C2-correlator} shows that $C_2(\Delta\mathbf R)$ remains finite and saturates to a plateau for $\Delta\mathbf R \gg \xi_0$, thereby demonstrating off-diagonal long-range order in the attractive Fermi gas.

\section{Measurement of different observable}
\label{app:observable-meas}
In this appendix we present the explicit Monte Carlo estimators used to evaluate the key observables discussed in the main text. 
Although these quantities can be written compactly in second quantization, their evaluation within a variational Monte Carlo framework requires reformulation in first-quantized form together with suitable importance-sampling strategies. 
We provide these derivations here both for completeness and to clarify the normalization conventions and numerical stabilizations employed in practice. 
Specifically, we describe: (i) the momentum-space occupation number $n(\mathbf k)$, which characterizes the distribution of fermions in reciprocal space; and (ii) the two-body reduced density matrix (2-RDM), whose eigenspectrum directly diagnoses off-diagonal long-range order. 
Each subsection below details the corresponding estimator and the Monte Carlo procedure used in its evaluation.



\subsection{Momentum-space occupation number}
The momentum-space occupation number can be written in first-quantized form as
\begin{align}
    \langle \hat n(\mathbf k)\rangle
     = \frac{1}{\mathcal V\,\mathcal Z}\!
        \int d\mathbf x'_1\, d\mathbf x'_2\, d\tilde{\mathbf R}\,
        e^{i\mathbf k\cdot(\mathbf x'_1-\mathbf x'_2)}\,
        \Psi^{*}(\mathbf x'_1,\tilde{\mathbf R})\,
        \Psi(\mathbf x'_2,\tilde{\mathbf R}),
\end{align}
where \(\tilde{\mathbf R}=(\mathbf x_2,\ldots,\mathbf x_N)\), \(\mathcal V\) is the system volume, and
\(\mathcal Z=\int d\mathbf X\,|\Psi(\mathbf X)|^2\).

To evaluate this via Monte Carlo, we use the importance-sampling density
\begin{align}
    p(\mathbf x'_1,\mathbf x'_2,\tilde{\mathbf R})
    = \frac{1}{\mathcal N}\,
      \big|\Psi(\mathbf x'_1,\tilde{\mathbf R})\,\Psi(\mathbf x'_2,\tilde{\mathbf R})\big|,
    \qquad
    \mathcal N
    = \int d\mathbf x'_1\, d\mathbf x'_2\, d\tilde{\mathbf R}\,
      \big|\Psi(\mathbf x'_1,\tilde{\mathbf R})\,\Psi(\mathbf x'_2,\tilde{\mathbf R})\big|.
\end{align}
Define the relative phase
\begin{align}
    \theta(\mathbf x'_1,\mathbf x'_2,\tilde{\mathbf R})
    = \mathrm{Arg}\!\left[\Psi^{*}(\mathbf x'_1,\tilde{\mathbf R})\,\Psi(\mathbf x'_2,\tilde{\mathbf R})\right].
\end{align}
Then
\begin{align}
    \langle \hat n(\mathbf k)\rangle
    = \frac{\mathcal N}{\mathcal V\,\mathcal Z}\;
      \mathbb E_{\tilde{\mathbf X}\sim p}\!\left[
        e^{i\mathbf k\cdot(\mathbf x'_1-\mathbf x'_2)}\,
        e^{i\theta(\mathbf x'_1,\mathbf x'_2,\tilde{\mathbf R})}
      \right].
\end{align}

The overall factor \(\mathcal N/(\mathcal V\,\mathcal Z)\) can be fixed by particle-number conservation,
\(\sum_{\mathbf k} \langle \hat n(\mathbf k)\rangle=N\).
In practice we compute the unnormalized expectation and rescale so that this sum rule is satisfied.

\subsection{Two-body reduced density matrix}
The zero–center-of-mass ($\mathbf Q=0$) 2-RDM in momentum space is
\begin{align}
    \Gamma_{\mathbf k,\mathbf k'}=\big\langle \hat \Delta^\dagger(\mathbf k)\,\hat \Delta(\mathbf k')\big\rangle,
    \qquad \hat \Delta(\mathbf k)=\hat c_{-\mathbf k}\hat c_{\mathbf k}.
\end{align}
In first quantization this can be written as
\begin{align}
    \Gamma_{\mathbf k,\mathbf k'}
    = \frac{1}{\mathcal V^2\,\mathcal Z}
      \!\int\! d\tilde{\mathbf R}\, d\mathbf x_1 d\mathbf x_2 d\mathbf x'_1 d\mathbf x'_2\;
      \Psi^*(\mathbf x_1,\mathbf x_2,\tilde{\mathbf R})\,
      \Psi(\mathbf x'_1,\mathbf x'_2,\tilde{\mathbf R})\,
      e^{-i\mathbf k\cdot(\mathbf x_1-\mathbf x_2) + i\mathbf k'\cdot(\mathbf x'_1-\mathbf x'_2)},
\end{align}
where $\tilde{\mathbf R}=(\mathbf x_3,\ldots,\mathbf x_N)$, $\mathcal Z=\int d\mathbf X\,|\Psi(\mathbf X)|^2$, and $\mathcal V$ is the spatial volume of integration.

To evaluate $\Gamma_{\mathbf k,\mathbf k'}$ by Monte Carlo, define the importance sampling density
\begin{align}
    p(\tilde{\mathbf X})
    = \frac{1}{\mathcal N}\;
      \big|\Psi(\mathbf x_1,\mathbf x_2,\tilde{\mathbf R})\,\Psi(\mathbf x'_1,\mathbf x'_2,\tilde{\mathbf R})\big|,
    \qquad
    \tilde{\mathbf X}=(\mathbf x_1,\mathbf x_2,\mathbf x'_1,\mathbf x'_2,\tilde{\mathbf R}),
\end{align}
normalized by $\int d\tilde{\mathbf X}\,p(\tilde{\mathbf X})=1$. Writing
\[
\theta(\tilde{\mathbf X})
=\text{Arg}\,\Big[\Psi^*(\mathbf x_1,\mathbf x_2,\tilde{\mathbf R})\,
        \Psi(\mathbf x'_1,\mathbf x'_2,\tilde{\mathbf R})\Big],
\]
yields the unbiased estimator
\begin{align}
\label{eq:two-rdm-estimator}
    \Gamma_{\mathbf k,\mathbf k'}
    = \frac{\mathcal N}{\mathcal V^2\,\mathcal Z}\;
      \mathbb{E}_{\tilde{\mathbf X}\sim p}\!\left[
      e^{i\theta(\tilde{\mathbf X})}\,
      e^{-i\mathbf k\cdot(\mathbf x_1-\mathbf x_2)+i\mathbf k'\cdot(\mathbf x'_1-\mathbf x'_2)}
      \right].
\end{align}
The unknown prefactor can be obtained from the same samples via
\begin{align}
    \frac{\mathcal N}{\mathcal V^2\,\mathcal Z}
    = \left(
        \mathbb{E}_{\tilde{\mathbf X}\sim p}
        \left[\left|\frac{\Psi(\mathbf x_1,\mathbf x_2,\tilde{\mathbf R})}
                         {\Psi(\mathbf x'_1,\mathbf x'_2,\tilde{\mathbf R})}\right|\right]
      \right)^{-1},
\end{align}
since $\mathbb{E}_{p}\!\left[ \big|\Psi/\Psi'\big| \right]=(\mathcal Z\,\mathcal V^2)/\mathcal N$. For numerical stability we evaluate $e^{i\theta}$ from complex log-amplitudes and average the estimator over the swap $(\mathbf x_1,\mathbf x_2)\!\leftrightarrow\!(\mathbf x'_1,\mathbf x'_2)$.

\emph{Symmetries and post-processing.}
By construction, $\Gamma$ is Hermitian and positive semidefinite:
\begin{align}
    \Gamma_{\mathbf k',\mathbf k}=\Gamma_{\mathbf k,\mathbf k'}^*,\qquad
    \sum_{\mathbf k,\mathbf k'}f^*_{\mathbf k}\Gamma_{\mathbf k,\mathbf k'}f_{\mathbf k'} \ge 0
    \quad \forall\,\{f_{\mathbf k}\}.
\end{align}
For spinless fermions $\hat\Delta(-\mathbf k)=-\hat\Delta(\mathbf k)$, which implies
\begin{align}
    \Gamma_{-\mathbf k,\mathbf k'}=-\,\Gamma_{\mathbf k,\mathbf k'},\qquad
    \Gamma_{\mathbf k,-\mathbf k'}=-\,\Gamma_{\mathbf k,\mathbf k'},\qquad
    \Gamma_{-\mathbf k,-\mathbf k'}=\Gamma_{\mathbf k,\mathbf k'}.
\end{align}
We enforce these relations by symmetrization,
\begin{align}
    \Gamma_{\mathbf k,\mathbf k'}
    \leftarrow \tfrac14\!\left(
       \Gamma_{\mathbf k,\mathbf k'}-\Gamma_{-\mathbf k,\mathbf k'}-\Gamma_{\mathbf k,-\mathbf k'}+\Gamma_{-\mathbf k,-\mathbf k'}
      \right),
    \qquad
    \Gamma \leftarrow \tfrac12\!\left(\Gamma+{\Gamma}^{\dagger}\right).
\end{align}
Because the estimator in Eq.~\eqref{eq:two-rdm-estimator} is generally non-positive, sampling noise can produce a cluster of small spurious eigenvalues in an interval $[-\varepsilon,\varepsilon]$, where $\varepsilon\to 0$ as the number of sample increases. After enforcing the symmetry relations above, Monte Carlo noise typically produces a continuum band of small eigenvalues around zero. We remove this band by \emph{magnitude} thresholding: diagonalize $\Gamma=U\Lambda U^\dagger$ and set
\[
\Lambda'_{ii}=\begin{cases}
0,& |\Lambda_{ii}|<\epsilon,\\
\Lambda_{ii},& \text{otherwise},
\end{cases}
\qquad
\Gamma_\epsilon=U\Lambda' U^\dagger.
\]
The scale $\varepsilon$ is chosen from the noise bulk (e.g., by a gap heuristic or the percentile of $|\lambda_i|$ within the small-eigenvalue cluster). This procedure projects out noise-dominated modes while preserving detached, physically meaningful eigenvalues (e.g., the condensate mode). 

\begin{table*}[t]
    \centering
    \caption{\justifying Number of Monte Carlo samples $N_{\mathrm{s}}$ used to estimate the observables reported in the main text and Supplementary Material. For each observable, we performed 300 burn-in steps, and here $N_{\mathrm{s}}$ counts post burn-in configurations per Markov chain.}
    \begin{tabular*}{\linewidth}{l l}
        \hline\hline
        Observable &  \hspace{4cm} Samples $N_{\mathrm{s}}$ \\[2pt]
        \hline
        Inference energy $E(N)$ and variance 
            &  \hspace{4cm} $N_{\mathrm{s}}^{(E)} = 2^{16}$ \\[2pt]
        Pair-binding energy $E_{B}(N)$ 
            &  \hspace{4cm} $N_{\mathrm{s}}^{(E_{\mathrm{B}})} = 2^{16}$ \\[2pt]
        $C_{4}$-projected energies and overlaps 
        $|\langle \tilde{\Psi}_{m} | \Psi \rangle|^{2}$ 
            &  \hspace{4cm} $N_{\mathrm{s}}^{(C_{4})} = 2^{16}$ \\[2pt]
        Momentum occupation $n(\mathbf{k})$ 
            &  \hspace{4cm} $N_{\mathrm{s}}^{(n(\mathbf k))} = 2^{24}$ \\[2pt]
        Superconducting order parameter $\langle c_{-\mathbf k}c_{\mathbf k}\rangle$ & \hspace{4cm}  $N_s^{(\Delta)}=2^{24}$ \\
        Two-body RDM $\Gamma_{k,k'}$ 
            & \hspace{4cm} $N_{\mathrm{s}}^{(\Gamma)} = 2^{26}$ \\[2pt]
        Real-space pair correlators 
        $C_{1}(\delta\mathbf{r})$,  
            & \hspace{4cm} $N_{\mathrm{s}}^{(C_1)} = 2^{28}$ \\[2pt]
        Real-space pair correlators 
        $C_{2}(\Delta\mathbf{R})$, &  \hspace{4cm} $N_{\mathrm{s}}^{(C_2)} = 2^{26}$ \\
        \hline\hline
    \end{tabular*}
    \label{table:mc_samples}
\end{table*}

\begin{figure*}[t]
    \centering
    \includegraphics[width=0.5\linewidth]{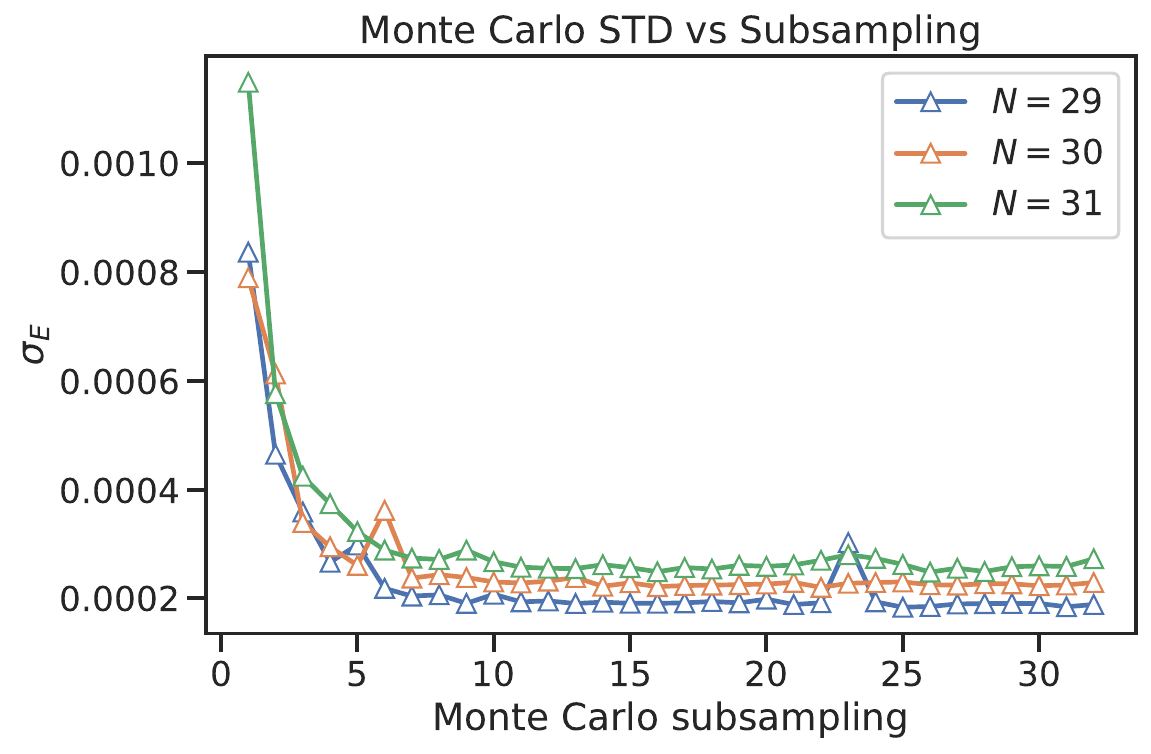}
    \caption{\justifying
    Standard deviation of the sample mean as a function of the subsampling interval. The blue, orange, and green curves correspond to $N=29, 30, 31$ with $U=-10.0$ and $L=41.798$.}
    \label{fig:std-subsampling}
\end{figure*}

\textit{Number of samples for different observables.}
For reproducibility, Table~\ref{table:mc_samples} lists the number of Monte Carlo samples used to estimate each observable.
For the energy estimates, we compute the integrated autocorrelation time $\tau_{\mathrm{int}}$ (Fig.~\ref{fig:std-subsampling}) to assess the quality of the MCMC sampling.
The integrated autocorrelation time quantifies how temporal correlations inflate the variance of the sample mean:
$\mathrm{Var}(\bar E)\approx (\sigma_E^2/N)\,(2\tau_{\mathrm{int}})$, so the effective sample size is
$N_{\mathrm{eff}}\approx N/(2\tau_{\mathrm{int}})$.
We vary the subsampling interval (the number of MC steps between consecutive recorded samples) and observe that the estimated standard error of the mean decreases and then plateaus for subsampling $\gtrsim 10$.
For the $N=29,30,31$ cases we obtain $\tau_{\mathrm{int}}\approx 9.847,\,5.943,\,8.887$ (in units of MC steps), indicating that choosing a subsampling interval comparable to or larger than this correlation scale makes successive recorded samples only weakly correlated and yields an effective sample size close to the nominal sample size.

\end{document}